\documentclass[amsmath,amsfonts,nofootinbib]{revtex4-1}

\usepackage{graphicx}
\usepackage{color}
\usepackage{physics}
\usepackage{amsmath}
\usepackage{array,mathtools,amssymb,booktabs}
\usepackage{bbold}
\usepackage{siunitx}
\usepackage[svgnames]{xcolor}
\usepackage{tabu}
\definecolor{linkblue}{HTML}{0066cc}
\usepackage[
    unicode,
    colorlinks=true,
    linkcolor=MidnightBlue,
    citecolor=Green,
    urlcolor=linkblue,
]{hyperref}
\usepackage{tikz}
\usetikzlibrary{positioning}
\usepackage{cleveref}

\newcommand{\vareps}{\varepsilon}

\AtBeginDocument{%
  \heavyrulewidth=.08em
  \lightrulewidth=.05em
  \cmidrulewidth=.03em
  \belowrulesep=.65ex
  \belowbottomsep=0pt
  \aboverulesep=.4ex
  \abovetopsep=0pt
  \cmidrulesep=\doublerulesep
  \cmidrulekern=.5em
  \defaultaddspace=.5em
}

\usepackage{soul}

\def\ket#1{| #1 \ra }
\def\bra#1{\la #1}

\def\la{\langle}
\def\ra{\rangle}

\def\beq{\begin{equation}}
\def\eeq{\end{equation}}
\def\bea{\begin{eqnarray}}
\def\eea{\end{eqnarray}}
\def\barr{\begin{array}}
\def\earr{\end{array}}

\def\Re{\mathop{\mbox{Re}}}

\begin{document}

\begin{titlepage}
\begin{flushright}
\end{flushright}
\vskip 0.5cm
\begin{center}
{\Large \bf Lattice investigation of custodial two-Higgs-doublet model\\at weak quartic couplings\\}
\vskip1cm {\large\bf
Guilherme~Catumba$^{a,b}$,
Atsuki~Hiraguchi$^{c,d}$,
Wei-Shu~Hou$^{e}$,
Karl~Jansen$^{f}$,
Ying-Jer~Kao$^{e,g}$,
C.-J.~David~Lin$^{c,h}$,
Alberto~Ramos$^{b}$,
Mugdha~Sarkar$^{c,e}$}\\ \vspace{.5cm}

\affiliation{}

{\normalsize {\sl
$^a$ Department of Physics, University of Milano-Bicocca \& INFN Milano–Bicocca, Piazza della Scienza 3, 20126, Milano, Italy\\
$^b$ Instituto de Fisica Corpuscular (IFIC), CSIC-Universitat de Valencia, 46071, Valencia, Spain\\
$^c$ Institute of Physics, National Yang Ming Chiao Tung
    University, Hsinchu 30010, Taiwan\\
$^d$ CCSE, Japan Atomic Energy Agency, 178-4-4, Wakashiba, Kashiwa, Chiba 277-0871, Japan\\
$^e$ Department of Physics, National Taiwan University, Taipei 10617, Taiwan\\
$^f$ Deutsches Elektronen-Synchrotron DESY, Platanenallee 6, 15738 Zeuthen, Germany\\
$^g$ Centre for Theoretical Physics and Centre for Quantum Science and Technology, National Taiwan University, Taipei 10607, Taiwan\\
$^h$ Centre for High Energy Physics, Chung-Yuan Christian University, Taoyuan 32023, Taiwan
    }}
\end{center}
\vskip1.0cm
\begin{center}
{\large\bf Abstract\\[10pt]} \parbox[t]{\textwidth}{{
The $SU(2){-}$gauged custodial two-Higgs-doublet model, which shares the same global-symmetry properties with the standard model, is studied
non-perturbatively on the lattice. The additional Higgs doublet enlarges the
scalar spectrum and opens the possibility for spontaneous
breaking of the global symmetry.
In this work we start by showing the occurrence of spontaneous breaking of the custodial symmetry in a region of the parameter space of the model.
Following this, both the spectrum and the running of the gauge coupling of
are examined at weak quartic couplings in the presence of the custodial symmetry.  The calculations are performed with energy cutoffs ranging from 300 to 600 GeV
on a line of constant standard model physics, obtained by tuning bare couplings to fix the ratio between the masses of the Higgs and the $W$ bosons, as well as the value of the renormalized gauge coupling at the scale of the $W$ boson mass.   
The realizable masses for the additional scalar
states are explored. For the choice of bare quartic couplings in this work, the
estimated lower bound of these masses is found to be well below the $W$ boson mass, and independent
of the cutoff.  We also study the finite temperature electroweak transition along this line of constant standard model physics, revealing properties of a smooth crossover behavior.
  }}
\end{center}

\vskip0.5cm
\end{titlepage}

\tableofcontents
\newpage
\section{Introduction}
\label{sec:intro}

The standard model (SM) has been very successful in describing
experimental results at the Large Hadron Collider (LHC) and other
facilities.  Nevertheless, physics beyond the SM (BSM) is needed in order to account for
many phenomena.  For instance, there are no dark-matter candidates in
the theory.  Another prominent problem is the lack of explanation for
the excess of matter over antimatter in the universe.  As suggested by
Sakharov~\cite{Sakharov_1991}, to explain electroweak
baryogenesis CP violation and a strong
first-order electroweak phase transition (EWPT) is needed.  The amount of CP
violation present in the SM is not enough for this purpose. In
addition, it is well established that the SM does not contain a
first-order electroweak phase transition at the measured Higgs boson
mass, $m_{h} \approx 125$
GeV~\cite{Kajantie_1996,Evertz:1986af,Rummukainen_1998,Laine_2013,D_Onofrio_2016}.
Such issues have been driving the search for BSM physics.

One natural scenario of BSM physics is the existence of extra scalar
fields.  While the Higgs sector of the SM is the minimal structure for
realising electroweak symmetry breaking, there are no compelling
reasons for the absence of additional Higgs doublets.  Indeed, a
structure analogous to fermion families can be introduced in the
scalar sector.  Such an extension of the SM can, arguably, result in
enough CP violation, and it can also alter the character of the
electroweak phase transition.  These features motivate the search of BSM Higgs doublets.  Taking into
account the phenomenological constraints imposed by the  measurement
of the $\rho$ parameter \cite{Workman:2022ynf}, a viable and simple
extension of the SM is the addition of a second $SU(2)$ doublet of
scalar fields.  This is known as the two-Higgs-doublet model (2HDM).

In the continuum, we denote the fields of the two scalar doublets by ($i = 1,2$)
\begin{equation}
\label{eq:cont_doublets}
\Phi^{c}_{i} = \left ( \begin{array}{c} \phi_{i}^{1} + i\phi_{i}^{2}\\ \\\phi_{i}^{3} + i \phi_{i}^{4} \end{array} \right ) \, .
\end{equation}
The Lagrangian of the 2HDM, coupled to $SU(2)$ gauge fields is 
\begin{equation}
  \mathcal L = \sum_{i=1}^{2} \left(D_{\mu}\Phi^{c}_{n}\right)^{\dagger}\left(D^{\mu}\Phi^{c}_{n}\right) - \frac{1}{4}G_{\mu\nu}^{a} G^{a,\mu\nu} - V_{\text{2HDM}}(\Phi^{c}_{1},\Phi^{c}_{2}) \, ,
\end{equation}
where $D_{\mu}= \partial_{\mu} - igA_{\mu}^{a}\sigma^{a}/2$
is the covariant derivative, with $A_{\mu}^{a}$ the real components of
the $SU(2)$ algebra-valued gauge field, $\sigma^{a}$ the Pauli
matrices, $g$ the gauge coupling constant, and
$G_{\mu\nu}^{a}= \partial_{\mu}A_{\nu}^{a} - \partial_{\nu}A_{\mu}^{a} + g\varepsilon^{abc}A_{\mu}^{b}A_{\nu}^{c}$
the field strength tensor.
The most general gauge-invariant scalar potential, $V_{\text{2HDM}}(\Phi^{c}_{1},\Phi^{c}_{2})$, can be written in
terms of the products $\Phi_{i}^{c,\dagger}\Phi^{c}_{j},~i,j=1,2$.
Since we are interested in the lattice application, we consider from
now on the case of real couplings only. This is the CP-conserving 2HDM
\cite{Haber_2011,ONeil:2009fty}.  With this choice, the most general
renormalizable scalar potential of the 2HDM has 10 real parameters,
and reads
\begin{equation}
  \label{eq:cont_potential}
\begin{aligned}
  V_{\text{2HDM}} &= \mu_{11}^{2}\Phi_{1}^{c,\dagger}\Phi^{c}_{1} + \mu_{22}^{2}\Phi_{2}^{c,\dagger}\Phi^{c}_{2} + \mu_{12}^{2}\Re (\Phi_{1}^{c,\dagger}\Phi^{c}_{2}) + \lambda_1 (\Phi_{1}^{c,\dagger}\Phi^{c}_{1})^2 + \lambda_2 (\Phi_{2}^{c,\dagger}\Phi^{c}_{2})\\
  &+ \lambda_3(\Phi_{1}^{c,\dagger}\Phi^{c}_{1})(\Phi_{2}^{c,\dagger}\Phi^{c}_{2})
  + \lambda_4(\Phi_{1}^{c,\dagger}\Phi^{c}_{2})(\Phi_{2}^{c,\dagger}\Phi^{c}_{1}) + \lambda_{5} \Re(\Phi_{1}^{c,\dagger}\Phi^{c}_{2})^{2}\\
  &  +\Re(\Phi_{1}^{c,\dagger}\Phi^{c}_{2})\left[ \lambda_{6} (\Phi_{1}^{c,\dagger}\Phi^{c}_{1}) + \lambda_{7} (\Phi_{2}^{c,\dagger}\Phi^{c}_{2})\right] \, .
\end{aligned}
\end{equation}

The above model yields an enlarged spectrum compared to the SM Higgs sector.  Its details depend crucially on the region of the parameter space.
In the perturbative treatment of the $SU(2){-}$gauged scalar theory with the potential in eq.~(\ref{eq:cont_potential}), the relevant parameter regions are those with the Higgs mechanism active, where the spectrum is expected to
contain five scalar states,
$h,H, A, H^{\pm}$.  In the Higgs-basis, where only one of the fields
acquires a vacuum expectation value, $v$ ---$\phi_{2}$ in the case at
hand---, the analysis of the mass matrices reveals the following
tree-level relations \cite{Haber:2006ue},
\begin{align}
  \label{eq:tl_mass_Hpm}
    &m_{H^\pm}^2 = \mu_{11}^{2} + \frac{1}{2}\lambda_3 v^2 \, ,\\
  \label{eq:tl_mass_a}
    &m_{A}^2 = m_{H^{\pm}}^{2} + \frac{1}{2}[\lambda_{4} - \lambda_{5}]v^2 \, ,\\
  \label{eq:tl_mass_matrix}
    &\mathcal M = v^2\mqty(\mu_{22}^{2}/v^{2} && \lambda_6  \\
                 \lambda_6 && \mu_{11}^{2}/v^2 + \frac{1}{2}[\lambda_{3} + \lambda_{4} + \lambda_{5}]
                 )\, .
\end{align}
The `charged' scalar states labelled by $H^{\pm}$ and the pseudo-scalar state, $A$,
have well defined masses. On the other hand, the remaining two CP-even
scalars require the diagonalization of $\mathcal M$.  However, in the
$\mathbb Z_{2}$-symmetric limit with
$\mu_{12}=\lambda_{6}=\lambda_{7}=0$, it is straightforward to obtain
$m_{h}^{2} = v^{2}\mu_{22}^{2}$ and
$m_{H}^{2} = \mu_{11}^{2}/v^2 + \frac{1}{2}[\lambda_3 + \lambda_4 + \lambda_5]$.

Other than the enlarged spectrum, this extension can have a rich
phenomenology, including, as we have mentioned, extra CP violation and possibly a first order
electroweak phase transition.  Additionally, these models can also
have other phenomenological applications.  In the class of models discussed in refs.~\cite{Ma:2006km,Barbieri_2006,honorez_inert_2007} the extra
states in the spectrum are used as dark matter candidates, and are
modeled to not interact with the quarks and leptons.  

While the augmented parameter space of the extended theory is a
desirable feature due to the rich phenomenology it can provide, the
parameters of the potential cannot be chosen in a completely arbitrary
way since, for instance, there are requirements of stability and
boundedness of the potential
\cite{maniatis_stability_2006,ivanov_algorithmic_2018}.  Furthermore,
any proposed 2HDM model should comply with current experimental
evidence for SM quantities.  In particular, it should suppress the flavour-changing neutral current (FCNC) processes,
as constrained by experiments.  On the other hand,
various studies have shown that the phenomenology of the theory is
dictated by the symmetries of the scalar potential
\cite{PhysRevD.75.035001,Pilaftsis:2011ed,Grzadkowski:2014ada,Alves:2018kjr,PhysRevD.77.015017},
particularly, in ref. \cite{ferreira_symmetries_2021} the symmetries
were related to physical parameters.  One of the earliest
symmetry-related proposals \cite{Glashow:1976nt,Paschos:1976ay} to
eliminate tree-level FCNC  takes the Lagrangian to be
$\mathbb Z_{2}$-symmetric under the change of sign of one of the
scalar doublets.  In ref.~\cite{HOU1992179}, however, it was claimed that
the $\mathbb Z_{2}$-symmetry may not be required to suppress the FCNC processes,
if certain mass-mixing hierarchies are found in the fermion flavour
sector, with an additional alignment of the CP-even Higgs mass
eigenstates \cite{Hou_2018}. This in turn may correspond to
$\order{1}$ quartic couplings, that are also thought to be required for a
first-order electroweak phase transition \cite{Fromme:2006cm}.

The amount of literature on the 2HDM is vast and ongoing (a general
review can  be found in ref.~\cite{branco_theory_2012}). There are analyses
of the parameter space, stability and boundedness of the potential~\cite{Sher:1988mj,maniatis_stability_2006,ivanov_algorithmic_2018},
phenomenological implications~\cite{PhysRevD.75.035001,Pilaftsis:2011ed,Grzadkowski:2014ada,Alves:2018kjr,PhysRevD.77.015017},
and group theoretical aspect of the global symmetries~\cite{ferreira_symmetries_2021}.  Most of these studies are performed
at tree-level, and while the scalar sector of the SM is weakly
coupled, recent investigations suggest that large scalar-field quartic
couplings in the BSM sector may be needed to have a strong first order
EWPT~\cite{Fromme:2006cm,dorsch_strong_2013,basler_strong_2017,bernon_new_2018}.
The large values of quartic coupling in itself puts in doubt the
perturbative approach to consider these questions
(cf.~\cite{Kainulainen:2019kyp}).  Moreover, scalar field theories are
expected to be trivial~\cite{Luscher:1987ay,Luscher:1987ek,Luscher:1988uq, Luscher:1988gc, Aizenman:2019yuo}.  They must be understood as
effective theories, only valid in the weakly coupled regime.  In summary,
the large quadratic couplings needed in many phenomenological models
and the triviality of scalar field theories requires a study beyond
perturbation theory.

Lattice techniques allow a fully non-perturbative study of scalar
models.  In this work we consider a simplified version of the Higgs
theories without the photon and the fermions, namely a $SU(2)$-Higgs
model. This theory describes the interaction of the two complex
doublets in the fundamental representation with the $SU(2)$ gauge
fields.  Unlike perturbative investigations, there have not been many
lattice studies of 2HDM.  In refs.~\cite{wurtz_effect_2009,
lewis_spontaneous_2010} the phase structure and the presence of
spontaneous symmetry breaking on the 2HDM was studied, while
\cite{maas2014observables} presented exploratory results on the
running of the gauge coupling in different sectors of the parameter
space.  Furthermore, among other BSM theories, the case of the
addition of an extra singlet to the Lagrangian has been explored also
on the lattice \cite{Niemi:2024axp}.  Nevertheless, there has not been
any lattice computation for the spectrum of the 2HDM. Additionally,
the only existing finite temperature investigations of the model were
carried out using a 3-dimensional effective field theory
\cite{Laine:2000rm, Kainulainen:2019kyp}.  In this work we tackle both
of these questions in the weakly coupled 2HDM in four dimensions.  Our
calculaion is the first lattice study of the 2HDM on a line of
constant SM physics.  This line is obtained by tuning the bare
couplings to reproduce the physical values of the ratio between the SM
Higgs and the $W$ boson masses, $m_{h}/m_{W}$, as well as the
renormalized gauge coupling at the scale of $m_{W}$.  We investigate
the BSM scalar spectrum and the finite temperature transition at
different values of the lattice spacing, corresponding to the cutoff
scale in the range of 300 to 600 GeV.  We also examine the pattern of
spontaneous breaking of the global symmetries.  Similar to the finding
from the lattice study of the $SU(2)\times SU(2)$-symmetric 2HDM in
ref.~\cite{lewis_spontaneous_2010}, we confirm the occurrence of the
spontaneous breaking of the custodial $SU(2)$ symmetry in 2HDM.

Although the calculation described in this paper is carried out at
weak quartic couplings, it is an important first step of our lattice
exploration of 2HDM.  In particular, it sets the ground for future
work at large renormalized quartic couplings on lines of constant SM
physics, which are highly relevant to the phenomenology of strong
first-order EWPT.  The preliminary results of this work can be found
in our contributions to the annual conferences on lattice field theory
~\cite{Catumba:2024oek, Catumba:2023jep,Sarkar:2022rti}.

\section{Standard model physics within the lattice 2HDM}
\label{sec:strategy}

One of the main objectives of this work is to build a line of constant
standard model physics within the lattice 2HDM at various values of lattice spacing, corresponding to different momentum cutoffs.  In order to do so we need to understand the phase structure
of the model, namely, the regions of parameter space where the theory is in the Higgs phase, and where and whether spontaneous breaking of the
global symmetries occurs.

In the following we introduce the lattice model, summarize the predictions about
the phase structure, and outline the strategy for tuning a line of constant SM
physics.

\subsection{Lattice Formulation}
\label{sec:strategy_lattice_formulation}

It is convenient to employ the quaternion formalism for lattice
simulations of scalar fields in the fundamental representation.  In
this approach, the two scalar doublets are represented by $2\times 2$ matrices ($i=1,2$),
\begin{equation}
\Phi_{i}(x) = \sum_{\alpha=1}^{4}\theta_{\alpha}\hat{\phi}_{i}^{\alpha}(x) \, ,
\end{equation}
where $\theta^{4}=\frac{1}{2}\mathbb{1}_{2\times2}$ and $~\theta^{k}=\frac{i}{2}\sigma^{k}$ for
$k=1,2,3$ with $\sigma^{k}$ being the Pauli matrices.  The real scalar fields on the lattice are related to those in eq.~(\ref{eq:cont_doublets}) through ($a$ is the lattice spacing)
\begin{equation}
\label{eq:latti_and_cont_phi}
 \hat{\phi}_{i}^{\alpha} = \frac{a}{\sqrt{\kappa_{i}}} \phi_{i}^{\alpha} \, ,
\end{equation}
with $\kappa_{i}$ related to the couplings in the continuum 2HDM potential in eq.~(\ref{eq:cont_potential}) through
\begin{equation}
a^{2} \mu^{2}_{ii} = \frac{1  - 8 \kappa_{i} - \lambda_{i} \kappa^{2}_{i}}{\kappa_{i}} \, .
\end{equation}

While the symmetry structure of the single Higgs model is
simple\footnote{In the scalar sector of the SM, the scalar potential
is invariant under a single $O(4)\simeq SU(2)\times SU(2)$.  Part of
this group is gauged, with the remaining $SU(2)$ being the only global
custodial symmetry.}, the global symmetries of the 2HDM can be more
complicated due to the enlarged parameter space. In order to
understand the latter, it is worth  starting from the simplest
non-interacting scalar theory without gauge fields. This model is
invariant under two independent $O(4)$ transformations, one for each
Higgs doublet. Since $O(4)\sim SU(2)\times SU(2)$, when gauging the
theory, two of the $SU(2)$ (one from each $O(4)$) are identified with
each other. After this step there remains a local gauge symmetry,
$SU(2)_L$, and two independent global $SU(2)$ symmetries.   
that the scalar fields $\Phi_i$ are in the fundamental representation
of $O(4)\sim SU(2)\times SU(2)$, of which one of the subgroups is
gauged, $SU(2)_L$, while the other remains as a global symmetry of the
theory.  This local and global symmetry structure is well understood
in the matrix formulation, since the fields $\Phi_{i}$ transform under
the $SU(2)_{L}$ gauge group by a left multiplication, and under the
global $SU(2)$ by right multiplication,
$\Phi_{i}\rightarrow L(x)\Phi_{i} R_i$, where both $L(x)$ and $R_i$
belong to the fundamental representations of $SU(2)_{L}$ and $SU(2)$,
respectively. While gauge symmetry forces both $SU(2)_L$ subgroups to
be the same, the remaining global symmetry structure depends on the
choice of terms in the action.

Previous lattice studies
\cite{wurtz_effect_2009,lewis_spontaneous_2010,maas2014observables}
have considered an enlarged global symmetry, namely
$SU(2)\times SU(2)$, where both scalars can be transformed
independently, $R_1\neq R_2$. In this work we take the
$\mathbb Z_{2}$-breaking terms in \cref{eq:cont_potential} to vanish,
$\mu_{12}=\lambda_{6}= \lambda_{7} = 0$, defining the so-called inert
models. Moreover, by imposing the condition $\lambda_{4}=\lambda_{5}$
we obtain the custodial limit of the 2HDM, where the model shares the
same global symmetry with the SM.  That is, its  action is invariant
under global custodial $SU(2)$ transformations where both scalar
doublets transform simultaneously and identically, $R_1=R_2$.

Contrary to the single Higgs case, where no spontaneous symmetry
breaking can occur (the gauge symmetry, including its global part, as
well as the custodial $SU(2)$ are never broken), the enlarged
parameter space of the 2HDM allows for the breaking of the global
symmetries. In this work we study the occurrence of spontaneous
symmetry breaking of the custodial symmetry in the 2HDM. This study is
relevant for characterizing the different regions of the parameter
space, and in particular for finding which of these are suitable for
reproducing low energy SM physics.

The lattice action for the custodial 2HDM used in this work can then be written as ($U_{\mu}(x)$ is the link variable that is an element of the $SU(2)$ gauge group)
\begin{align}
  \label{eq:lattice action}
  S_{\text{2HDM}}= \sum_{x}&\bigg[\sum_{i=1}^{2} \bigg\{ \sum_{\mu=1}^{4} -2 \kappa_{i}\Tr \left( \Phi_{i}^{\dagger}U_{\mu}\Phi_{i}(x+\hat\mu) \right)+  \Tr \left( \Phi_{i}^{\dagger}\Phi_{i} \right) + \eta_{i}\left[ \Tr \left( \Phi_{i}^{\dagger}\Phi_{i} \right) - 1 \right]^{2}\bigg\}  \nonumber \\
  & + \eta_{3}\Tr \left( \Phi_{1}^{\dagger}\Phi_{1} \right)\Tr \left( \Phi_{2}^{\dagger}\Phi_{2} \right) + 2\eta_{4}\Tr \left( \Phi_{1}^{\dagger}\Phi_{2} \right)^{2}\bigg] + S_{\text{YM}} \, ,
\end{align}
where fields without argument are evaluated at the lattice point $x$,
and $S_{\text{YM}}$ is the standard Wilson plaquette action
\begin{equation}
  \label{eq:YM action}
  S_{\text{YM}} = \beta\sum_{x}\sum_{\mu>\nu}\left[ 1 - \frac{1}{2}\Re \Tr U_{\mu\nu}(x) \right] \, ,
\end{equation}
with $U_{\mu\nu}$ the plaquette on the $(\mu,\nu)$ plane and $\beta = 4/g^{2}$.  The bare
lattice quartic couplings in \cref{eq:lattice action} are related to their continuum
counterparts by 
\begin{eqnarray}
 \eta_{i} &=& \lambda_{i} \kappa_{i}^{2} \mbox{ } \mathrm{for}\mbox{ } i = 1,2\, ,\nonumber \\
 \eta_{i} &=& \lambda_{i} \kappa_{1} \kappa_{2} \mbox{ } \mathrm{for}\mbox{ } i = 3,4,5\, .
\end{eqnarray}

For the numerical simulations, the scalar and gauge field
configurations were generated using the Hybrid Monte Carlo (HMC)
algorithm (details in \cref{sec:HMC}).  The code implementation for
NVIDIA GPUs in the Julia programming language can be found in
\cite{lgpu_su2} (documentation can be found in \cite{lgpu}).  All
error analysis were done using the $\Gamma$-method
\cite{Wolff:2003sm}, using the free implementation available in
\cite{aderrors} (see also \cite{Ramos:2018vgu}).

\subsection{Phase structure, Order Parameters \& Symmetry breaking}
\label{sec:strategy_phase_structure}

While the single-Higgs-doublet model, governed by three bare
couplings, has a simple phase structure\footnote{In fact, only a
single phase exists in the Higgs-gauge theory. The confinement and
Higgs sectors are analytically connected.}, the enlarged parameter
space of the 2HDM, as well as the additional symmetries of the system,
give rise to a more complicated structure.

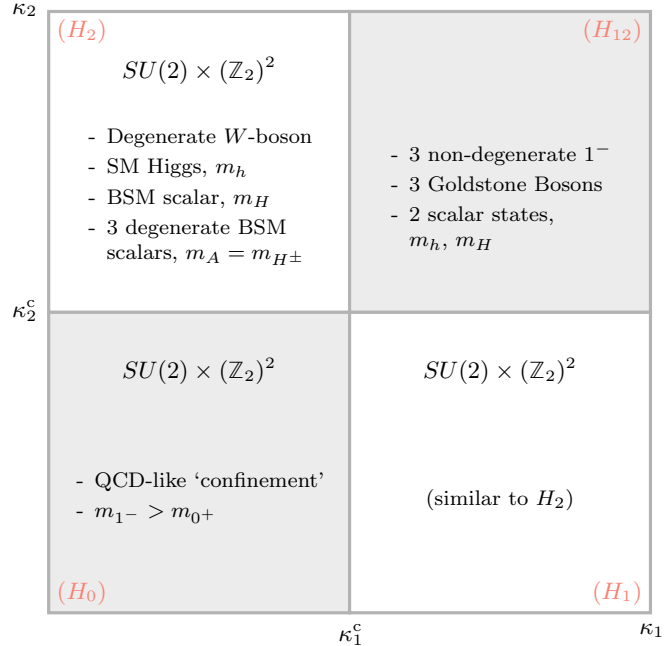
\begin{figure}[htb!]
  \centering
\begin{tikzpicture}[very thick]
    \definecolor{salmon}{RGB}{250, 128, 114}
    \def\size{4cm}
    \node[fill=lightgray!30,draw=black!30, minimum size=\size, inner sep=0pt] (H0) {};
    \node[fill=white,draw=black!30,minimum width=\size, minimum height=\size, inner sep=0pt, above=-\pgflinewidth of H0] (H2) {};
    \node[fill=white,draw=black!30,minimum width=\size, minimum height=\size, inner sep=0pt, right=-\pgflinewidth of H0] (H1) {};
    \node[fill=lightgray!30,draw=black!30, minimum size=\size, inner sep=0pt,  above right=-\pgflinewidth and -\pgflinewidth of H0] (H12) {};

    \node[anchor=east] at (H0.north west) {$\kappa_{2}^{\rm c}$};
    \node[anchor=east] at (H2.north west) {$\kappa_{2}$};
    \node[anchor=south west] at (H0.south west) {\textcolor{salmon}{$(H_{0})$}};
    \node[anchor=north west] at (H2.north west) {\textcolor{salmon}{$(H_{2})$}};
    \node[anchor=north] at (H0.south east) {$\kappa_{1}^{\rm c}$};
    \node[anchor=north] at (H1.south east) {$\kappa_{1}$};
    \node[anchor=south east] at (H1.south east) {\textcolor{salmon}{$(H_{1})$}};
    \node[anchor=north east] at (H12.north east) {\textcolor{salmon}{$(H_{12})$}};

    \node[yshift=-.5cm,anchor=north] at (H2.north) {$SU(2)\times (\mathbb Z_{2})^{2}$};
    \node[yshift=-.5cm,anchor=north] at (H1.north) {$SU(2)\times (\mathbb Z_{2})^{2}$};
    \node[yshift=-.5cm,anchor=north] at (H0.north) {$SU(2)\times (\mathbb Z_{2})^{2}$};

    \footnotesize
    \node[yshift=-.5cm,xshift=-.3cm,anchor=center] at (H0.center)
    {\begin{varwidth}{\linewidth}\begin{itemize}
        \setlength\itemsep{-0em}
        \item[-] QCD-like `confinement'
        \item[-] $m_{1^{-}}>m_{0^{+}}$
    \end{itemize}\end{varwidth}};

    \node[yshift=-.5cm,xshift=-.3cm,anchor=center] at (H2.center)
    {\begin{varwidth}{\linewidth}\begin{itemize}
        \setlength\itemsep{-0em}
        \item[-] Degenerate $W$-boson
        \item[-] SM Higgs, $m_{h}$
        \item[-] BSM scalar, $m_{H}$
        \item[-] 3 degenerate BSM\\ scalars, $m_{A}= m_{H^{\pm}}$
    \end{itemize}\end{varwidth}};

    \node[yshift=-.5cm,xshift=-.0cm,anchor=center] at (H1.center)
    {\centering (similar to $H_{2}$)};

    \node[yshift=-.5cm,xshift=-.3cm,anchor=center] at (H12.center)
    {\begin{varwidth}{\linewidth}\begin{itemize}
        \setlength\itemsep{-0em}
        \item[-] 3 non-degenerate $1^{-}$
        \item[-] 3 Goldstone Bosons
        \item[-] 2 scalar states,\\ $m_{h}$, $m_{H}$
    \end{itemize}\end{varwidth}};


\end{tikzpicture}
\caption{Summary of the cutodial 2HDM parameter space with the global
symmetries and spectrum content for each of the sectors $H_{0}$,
$H_{1}$, $H_{2}$, $H_{12}$. $1^{-}$, and $0^{+}$ represent a vector
and a scalar state, respectively.}
\label{fig:scheme_sectors}
\end{figure}

In the case of the inert models, it is possible to divide the parameter space in
four different regions
where none, one, or both scalar doublets are in the Higgs phase
\cite{branco_theory_2012}. In the perturbative formulation, using the vacuum
expectation values, $v_{i}$, for each Higgs field this corresponds to:
\begin{equation}
  \label{eq:sectors_definition}
\begin{aligned}
  &(H_{0}): v_{1}=v_{2}=0,\\
  &(H_{2}): v_{1}=0, v_{2}\neq0,\\
  &(H_{1}): v_{1}\neq0, v_{2}=0,\\
  &(H_{12}): v_{1}\neq0, v_{2}\neq0
\end{aligned}
\end{equation}
These sectors divide the $\kappa_{1}, \kappa_{2}$ plane into four
sectors defined by the critical values,
$\kappa_{i}^{c}$.\footnote{Notice that the value of
$\kappa_{i}^{c}$ depends on the remaining couplings of the theory.}
Note, however, that not all four can be seen as separate phases since
some of these transitions are crossovers for certain regimes of the
couplings.  The global symmetries, predicted particle content, and
qualitative aspects of each of the phases are summarized in
\cref{fig:scheme_sectors}.  The tree-level bounds for these regions
are defined in ref.~\cite{PhysRevD.18.2574}, and in
refs.~\cite{wurtz_effect_2009,lewis_spontaneous_2010} these sectors were
studied for the $SU(2)\times SU(2)$-symmetric inert potential
non-perturbatively on the lattice.

Sectors $(H_{1}),(H_{2})$, and $(H_{12})$ have the Higgs mechanism (or the Frohlich-Morchio-Strocchi (FMS) mechanism~\cite{Frohlich:1980gj, Frohlich:1981yi}) active.
However, as will be shown in \cref{sec:results_spontaneous_sym_breaking}, the
global custodial $SU(2)$ symmetry is spontaneously broken in $(H_{12})$.  The absence
of the custodial symmetry, and the presence of massless Goldstone bosons in the
spectrum exclude this phase from phenomenological considerations if one is
interested in embedding SM physics in the 2HDM.\footnote{In fact, in
order to reproduce the experimentally observed measurements of the SM,
the \textit{alignment} limit of the 2HDM (where one of the scalars
reproduces the SM Higgs) seems to be favoured \cite{bernon_new_2018}.}


Sectors $(H_{1})$ and $(H_{2})$ are equivalent, and for definitiveness
we will work with the latter.  This is defined by
$\kappa_{2}>\kappa_{2}^{c}$, $\kappa_{1}<\kappa_{1}^{c}$, with the
field $\Phi_{2}$ reproducing the SM-like Higgs.  At tree-level this
model predicts the existence of three degenerate scalar particles,
$m_{A}^{2} = m_{H^{\pm}}^{2}=\mu_{11}^{2}+\lambda_{3}v^{2}/2$ (see
\cref{eq:tl_mass_Hpm,eq:tl_mass_a}). Additionally, since the mass
matrix, \cref{eq:tl_mass_matrix}, is diagonal in the inert limit, we
identify the SM Higgs with $m_{h}^{2}=\mu_{22}^{2}v^{2}$, and an
additional scalar with mass
$m_{H}^{2}=m_{H^{\pm}}^{2} + \lambda_{4}v^{2}$.


The study of the phase structure on the lattice, namely the
identification of the different sectors from
\cref{fig:scheme_sectors}, requires a proxy for an order parameter.  The different regions are
easily identified by monitoring global observables, i.e., volume
averaged operators, such as the average plaquette,
\begin{equation}
  P = \frac{1}{V}\sum_{x,\mu,\nu}\Tr U_{\mu\nu} \, ,
\end{equation}
the squared Higgs-field length,
\begin{equation}
\rho_{i}^{2}=\frac{1}{V}\sum_{x}\det(\Phi_{i}(x)) \, ,
\label{eq:rho_def}
\end{equation}
or the gauge-invariant links
\begin{align}
  L_{\Phi_{ij}}^{a} =\frac{1}{8V}\sum_{x,\mu}\Tr{\Phi_{i}^{\dagger}(x)U_{\mu}(x)\Phi_{j}(x+\hat\mu)\theta^{a}},\label{eq:Lphi}\\
  L_{\alpha_{ij}}^{a} =\frac{1}{8V}\sum_{x,\mu}\Tr{\alpha_{i}^{\dagger}(x)U_{\mu}(x)\alpha_{j}(x+\hat\mu)\theta^{a}}, \label{eq:global_obs}
\end{align}
where $\alpha_{i}$ is the `angular' part of the quaternion Higgs field,
$\Phi_{i}=\rho_{i}\alpha_{i},~\rho_{i}\in\mathbb R,~\alpha_{i}\in SU(2)$.  In
order to detect the transitions between the sectors in
\cref{eq:sectors_definition} we mostly look at
$L_{\alpha_{i}}\equiv L_{\alpha_{ii}}^{4}$ since it is a bounded quantity for
all values of the couplings \cite{wurtz_effect_2009}, and it is the most
sensitive to the transitions \cite{PhysRevD.41.2573}.\footnote{The quantity
$\rho_{i}^{2}$ can be related to the number of scalar states, while the
gauge-invariant link can be understood as the energy density of the gauge-Higgs
interaction,
$$L_{\Phi_{ii}}^{4} =  \frac{1}{8V}\pdv{}{\kappa_{i}}\log Z.$$
In the Higgs phase, the gauge interaction is screened by the `condensation' of
the scalar field, making the gauge field configurations smoother, and thus
increasing the average gauge-invariant link.  Heuristically, using the usual
symmetry breaking perspective the `symmetric' phase can be understood as a
random state of the $SU(2)$ part of the scalar field, $\alpha$, while it is in
an ordered state in the `broken' region.  This explains why $L_{\alpha}$
increases across the two regions.}


Other than finding the regions where the Higgs effect is active, we
are also interested in probing the existence of spontaneous symmetry
breaking (SSB) of the additional global symmetries in the 2HDM.  In
order to confirm the occurrence of SSB we modify the partition
function by adding an extra term $\vareps \mathcal B$ to the action,
\begin{equation}
    Z_{\vareps} = \int D\phi(x) \mbox{ }\mathrm{exp}( -S[\phi] + \vareps \mathcal B ),
\end{equation}
where $\mathcal B$ is a term that explicitly breaks the symmetry in
question.  In the modified theory there is a preferred direction in
the group space, such that we get a non-zero value,
$\expval{\mathcal O}_{\vareps}\neq 0$, for any quantity $\mathcal O$
that is not invariant under this symmetry. The occurrence of spontaneous
symmetry breaking can then be probed by the order parameter
$\expval{\mathcal O}$ defined by the limiting procedure
\begin{equation}
  \label{eq:limit_ssb}
  \expval{\mathcal O} =\lim_{\vareps \rightarrow 0} \expval{\mathcal O}_{\vareps} =
\begin{dcases}
0, & \textrm{symmetry~intact} \\
v, & \textrm{spontaneous breaking}.
\end{dcases}
\end{equation}
In practice, due to the finite volume of the lattice, the definition of the
order parameter must be altered to
\begin{equation}
  \label{eq:limit_ssb_lattice}
  \expval{\mathcal O} = \lim_{\vareps \rightarrow 0} \lim_{V\rightarrow\infty}\expval{\mathcal O}_{\vareps},
\end{equation}
where the thermodynamic limit is taken before the $\vareps\rightarrow 0$ limit.

In \cref{sec:results_spontaneous_sym_breaking} this procedure is considered for
studying the spontaneous symmetry breaking of the custodial $SU(2)$ by studying the
quantity $L_{\alpha_{12}}^{3}$ with
\begin{equation}
  \label{eq:ssb_breaking_term}
  \mathcal B =  \Tr\left[\Phi_{1}^{\dagger}(x)\Phi_{2}(x)\theta^{3}\right].
\end{equation}

\subsection{Standard Model physics on the lattice 2HDM}
\label{sec:strategy_phase_structure_constant_physics}

In this work, we examine carefully the spectrum of the BSM scalar states in the 2HDM described in secs.~\ref{sec:strategy_lattice_formulation} and \ref{sec:strategy_phase_structure}, i.e., the model that shares the same custodial symmetry with the SM.  We are particularly interested in exploring how the spectrum of these BSM states depends on the cutoff scale.  In this procedure, one  must embed the SM in this model by tuning the bare parameters to reproduce SM physics.  
This introduces the notion of a line of partially constant physics (LPCP), on which we fix the values of two dimensionless, renormalized SM quantities, namely, 
the ratio of the Higgs to the $W$ boson\footnote{
What we call the $W$ bosons in this work are the physical gauge invariant vector states that transform as a triplet under the global $SU(2)$ custodial symmetry group. Their masses can be determined by spectroscopy studies on the lattice. While this is not to be confused with the usual $W$ boson from gauge-fixed perturbation theory (a triplet under gauge symmetry), assuming the validity of the FMS mechanism allows to identify the perturbative states with the physical states.},
\begin{equation}
  \label{eq:R}
  R \equiv \frac{m_{h}}{m_{W}} \, ,
\end{equation}
and the renormalized running gauge coupling $g_{R}^{2}(\mu)$.   The ratio, $R$,
can also be related at tree-level to the renormalized quartic coupling of the
corresponding `broken' scalar, $R^{2}\propto \lambda_{R}/g_{R}^{2}$
\cite{fodor_simulating_1994}.  In the following, we fix the $R$-ratio to be
close to the SM value, $R = 1.5$, while the renormalized gauge coupling is set
to its physical value at the scale of the $W$ boson mass,
$g_{R}^{2}(\mu=m_{W})\equiv 4\pi \alpha_{W}(\mu=m_{W})\sim 0.5$.  In other words, we scan the space of bare couplings in eqs.~(\ref{eq:lattice action}) and (\ref{eq:YM action}) to study possible values of the BSM scalar masses while keeping $R$ and $g_{R}^{2}(\mu=m_{W})$ constant.  Since no BSM particles have been found experimentally, it is not of interest to perform tune a complete line of constant physics for the 2HDM by also keeping the masses
of the BSM states fixed.

To implement the above strategy involving the LPCP, a scheme for extracting the renormalized gauge coupling has to be specified.  For this purpose, we resort to the gradient-flow scheme~\cite{Luscher_2010_flow}.  
In this approach,
the renormalized gauge coupling is defined through the action density,
\begin{equation}
  \label{eq:action_density}
  \expval{E(x,t)} = -\frac{1}{4}\expval{G_{\mu\nu}^{a}(x,t)G_{\mu\nu}^{a}(x,t)} \, ,
\end{equation}
where $G_{\mu\nu}(x,t)$ is the flowed gauge field strength tensor,
\begin{equation}
    G_{\mu\nu}(x,t) = \partial_\mu B_\nu(x,t) - \partial_\nu B_\mu(x,t) + \left[ B_\nu(x,t), B_\mu(x,t)  \right] \, ,
\end{equation}
with the flowed gauge potential, $B_{\mu}(x,t)$ satisfying the equation,
\begin{equation}
    \label{eq:gf_continuum_main}
    \dv{B_\mu(x,t)}{t} = D_\nu G_{\nu\mu}(x,t) \, ,
\end{equation}
with the initial condition $B_{\mu}(x,0)=A_\mu(x)$. The effect of the gradient flow
is to smear the gauge fields over a region of radius $\sqrt{8t}$. In this work, we use the Clover discretization of the field strength
tensor, and the Wilson plaquette action for the implementation of the lattice version of eq.~(\ref{eq:gf_continuum_main}). For more details, see ref.~\cite{ramos_yang-mills_2015} and references therein.

The action density, \cref{eq:action_density}, at the scale
$\mu=1/\sqrt{8t}$ is a renormalized quantity at finite flow times.  It
has been computed in perturbation theory~\cite{luscher_perturbative_2011} in terms of the renormalized coupling
$g$,
\begin{equation}
  \label{eq:action_density_ren_coupling}
  t^{2}\expval{E(t)}\eval_{t=1/(8\mu^{2})} = \frac{3(N^{2}-1)g^{2}(\mu)}{128\pi^{2}}\left(1 + c_{1}g^{2}(\mu) + \order{g^{2}}\right) \, ,
\end{equation}
with the $c_{1}$ being a finite, scheme dependent coefficient, and $N=2$.   Using
this relation  we define the gradient-flow renormalized gauge coupling, and specify its physical value at the scale of the $W$ boson mass by,
\begin{equation}
  \label{eq:GF_coupling}
    g_{GF}^{2}(\mu)\equiv \frac{128\pi^{2}}{9} \eval{t^{2}\expval{E(t)}}_{t=1/8\mu^{2}}, \, \, \,  g_{GF}^{2}(\mu = m_{W}) = 0.5 \, ,
\end{equation}
where $\expval{E(t)}$ is obtained from the Euclidean spacetime average of
$\expval{E(x,t)}$.  In our lattice calculation, this is obtained by tuning the bare parameters to reproduce the condition,
\begin{align}
  \label{eq:S}
  S\equiv \sqrt{8t_{0}}m_{W} = 1.0 \, ,
\end{align}
where the $t_{0}/a^{2}$ is the flow time at which $g^{2}_{GF}(\mu)$ takes the value of 0.5.

Adjusting the bare parameters to satisfy the condition $R=1.5$ and $S=1$ at different lattice spacings, a LPCP is mapped out. The
interest is then to explore the realizable spectrum of the BSM
scalar sector along this LPCP.

\subsection{Physical spectrum \& interpolators}
\label{sec:spectrum_interpolators}

In order to study the spectrum of the theory, and also to obtain the
first condition for SM physics within the lattice 2HDM, \cref{eq:R},
we consider zero momentum, gauge-invariant composite operators at
Euclidean time $x_{4}$,
\begin{align}
    &S_{ij}^{a}(x_{4}) = \sum_{\vec x} \Tr\left[\Phi_i^\dagger(x)\Phi_j(x)\theta^{\alpha}\right] \, ,\label{eq:Sij}\\
    &W_{ij,\mu}^a(x_{4}) = \sum_{\vec x} \Tr\left[\Phi_i^\dagger(x)U_{\mu}(x)\Phi_j(x+\hat\mu)\theta^{\alpha}\right] \, ,\label{eq:Wij}
\end{align}
that can be classified according to the representations of the global symmetry
group \cite{maas_gauge_2016} as well as of the lattice $H(4)$ hypercubic
symmetry group.  While at any finite lattice spacing it is possible to establish
a correspondence between the $H(4)$ representations and the continuum angular
momentum and parity quantum numbers, $J^{P}$, in this work we did not try to
resolve the different lattice representations, assuming small modifications for
the low-lying spectrum.  See refs.~\cite{Wurtz:2015zsz,PhysRevD.88.054510} for a
dedicated analysis in the Higgs theory.

At leading order in the lattice spacing $S_{ij}^{\alpha}$ couples to a
$J^{P}=0^{+}$ state for any combination of $i,j$, and $\alpha$ (with the exception
$S_{ii}^{k}=0$ due to the properties $\Phi_{i}^{\dagger} \Phi_{i} \propto \mathbb{1}_{2\times2}$ and
$\Tr\sigma^k = 0$).  On the other hand, the interpolators $W_{ij,\mu}^\alpha$
can be understood in terms of continuum fields by performing a classical
expansion in the lattice
spacing\footnote{$U_\mu(x) = \mathrm{exp}(ia A_\mu^b \sigma^b/2) = 1 + ia A_\mu^b \sigma^b/2 + \order{a^2}$
and $\Phi(x\pm\hat \mu) = \Phi(x) \pm a\partial_\mu \Phi(x)$.}
\begin{align}
\label{eq:W_ij_mu}
    W_{ij,\mu}^\alpha  =\sum_{\vec x}\frac{1}{2}\Tr\left[\theta^\alpha \left(\Phi_i^\dagger(x)\Phi_j(x) + \Phi_i^\dagger(x)\Phi_j(x) + a (D_\mu\Phi_i)^\dagger \Phi_j - a \Phi_i^\dagger D_\mu\Phi_j \right) + \order{a^{2}} \right] \, .
\end{align}
This interpolator sources states with different quantum numbers, depending on
the indices $i,j$, and $\alpha$.  For $i=j$ and $\alpha=4$ (recall that $\theta^{4}=\mathbb 1_{2\times 2}$),  the
$\order{a^{0}}$ component of $W^{4}_{ii,\mu}$ couples to a scalar state $J^P=0^+$, and is the leading contribution.
With $i=j$ and $\alpha=1,2,3$, the $\order{a^{0}}$ terms on the right-hand side of eq.~(\ref{eq:W_ij_mu}) vanish because 
$\Tr\sigma^k=0$, and $W^{\alpha=k=1,2,3}_{ii,\mu}$ overlaps with a
$J^P = 1^-$ state.  In this case, $\mu=1,2,3$ correspond to the $J=1$
spin components, and $k=1,2,3$ to the $SU(2)$ triplet representation.
Finally, when $i\not= j$, the 
$\order{a^{0}}$ component in $W_{i\neq j,\mu}^{\alpha}$ is always non-zero, and this interpolator primarily couples to four scalar states
labelled by $\alpha$.  From the above discussion, we can assign the $J^{P}$ quantum numbers in the following way,
\begin{align}
\label{eq:interpolators_generic}
  S_{ii}^{4}\, ,~W_{ii}^{4} \rightarrow 0^{+}\, , && W_{ij}^{\alpha} \rightarrow 0^{+}~(i\neq j)\, , && W_{ii}^{k} \rightarrow 1^{-}\, .
\end{align}
Additionally, since the $J=1$ spin components have the same mass, in the
simulations we average over the $\mu=1,2,3$ directions when computing two-point
functions.


While the classification of the operators according to the angular momentum
representations is straightforward, the connection to the physical spectrum is
more involved.  The reason is that the spectrum depends not only on the sector of
the theory but also on the symmetries of the scalar potential, which can give
rise to degenerate or non-degenerate scalar multiplets in the broken phase, and
even Goldstone bosons.\footnote{Notice that in the most general 2HDM there is
mixing in the scalar sector, such that a variational method would be required to
extract the complete spectrum. This is not the case in the inert limit. On the
other hand, mixing always occurs in the sector $(H_{12})$.}  In this work, we are mostly interested in studying the spectrum in sector $(H_{2})$, where both $S_{22}^4$ and $W_{22}^4$
yield the mass of the SM Higgs, $m_{h}$, while the vector operator
$W_{22}^{k}$ sources the degenerate $SU(2){-}$triplet states that represent the $W$ bosons.  The
remaining four BSM scalar states are sourced by $S_{12}^a$ or $W_{12}^a$.

In the $(H_{2})$ sector of the custodial 2HDM, the mixing amongst scalar states is absent. This allows
us to extract the ground-state energies by computing the effective
mass, $m_{\rm eff}$, from the large Euclidean time behavior of the
correlators
\begin{equation}
  \label{eq:correlator}
  C_{\mathcal O}(x_{4}) = \expval{\mathcal O(x_{4})\mathcal O(0)} = \sum_{n} \abs{ \bra n \hat{\mathcal O} \ket 0 }^{2} e^{-x_{4}\Delta E_{n}} \, ,
\end{equation}
where $\Delta E_{n}$ are the energies relative to the vacuum, and $\mathcal O$
one of the local interpolators defined in \cref{eq:Sij,eq:Wij}.
The correlation functions are computed with smeared interpolators.  For this we
consider gradient (Wilson) flow smearing on the gauge links.\footnote{Gradient
flow smearing includes smearing in the time direction, which breaks locality in
time and may introduce contamination from excited states in the mass extraction.
For this reason, the smearing radius was always taken to be very small,
$\sqrt{8t_{\rm smear}}\ll t_{\rm m}$, where $t_{\rm m}$ is the minimal time in
lattice units to measure the effective mass.}  Additionally, we perform smearing
for the scalar fields using the Laplacian operator  \cite{PhysRevD.80.054506}
\begin{align}
  \phi^{(n+1)}(x) &= (1 + r\nabla^{2})\phi^{(n)}(x) \\
                 &= \phi^{(n)}(x) + r\sum_{\mu=1}^{3}\left( \tilde U_{\mu}(x)\phi^{(n)}(x+\hat\mu) - 2\phi^{(n)}(x) + \tilde U_{\mu}^{\dagger}(x-\hat\mu)\phi^{(n)}(x-\hat\mu) \right) \, ,\nonumber
\end{align}
where $\tilde U$ is the smeared link and $r$ controls the smearing radius. This
step is iterated $N$ times, $\phi^{(1)}=\phi \longrightarrow \phi^{(N)}$,
corresponding to a smearing radius $\sigma_{x}\sim \sqrt{2Nr}$.

%
%

%

%
%

%
\section{Results and discussion}
\label{sec:results}

In this section, we present the results from our numerical study.  We
start by investigating the spontaneous breaking of the custodial symmetry
in phase $(H_{12})$ in sec.~\ref{sec:results_spontaneous_sym_breaking}.  The absence of custodial symmetry in the spectrum in this phase makes it uninteresting for high-energy physics phenomenology.  This
leaves sectors $(H_{1})$ and $(H_{2})$ available for constructing BSM models that embed SM physics.
We then consider the tuning for
the line of constant SM physics, the LPCP described in sec.~\ref{sec:strategy_phase_structure_constant_physics}, in the $(H_{2})$ sector. The results of this tuning are
shown in sec.~\ref{sec:results_tuning}.
Section~\ref{sec:results_gauge_coupling} gives details of our exploration for the running behaviour of
the renormalized gauge coupling along this LPCP.  In sec.~\ref{sec:results_spectrum} we report our
study of the BSM scalar-state spectrum.  Finally the finite temperature
transition is discussed in sec.~\ref{sec:results_finite_temp}.

Except for sec.~\ref{sec:results_spontaneous_sym_breaking}, all of the
numerical results shown are obtained in sector $(H_{2})$. The qualitative
structure of the space of parameters, \cref{fig:scheme_sectors}, is
common for all inert models, as known from our previous works
\cite{Sarkar:2022rti,Catumba:2025obz}.  A detailed study of the phase
strucure, the spectrum, and the dependence on the bare couplings, can
be found in \cite{Catumba:2023jep}.  For the purpose of this
investigation it suffices to monitor the order parameters defined in
sec.~\ref{sec:strategy_phase_structure}, as well as the mass
ratio of the lightest scalar and vector states associated with the
SM Higgs and the $W$ bosons, to guarantee that we are working in the desired
sector.

%
%

\subsection{Spontaneous breaking of custodial symmetry}
\label{sec:results_spontaneous_sym_breaking}
In this subsection we consider the breaking of the $SU(2)$
global custodial symmetry.  In principle, such an investigation can be carried out by identifying
massless Goldstone bosons in the sector $(H_{12})$.  This, however, requires a dedicated study of the spectrum in the thermodynamic limit, and is resource-demanding.  In this work, we employ a different approach to examine this symmetry breaking pattern by investigating the observable, $L_{\alpha_{12}}^{j},j=1,2,3$, defined in \cref{eq:global_obs}.  This observable is not invariant under the custodial $SU(2)$ transformation.  As reported below, through the study of $\langle L_{\alpha_{12}}^{j}\rangle$ across the boundary between $(H_{2})$ and $(H_{12})$ sectors, we are able to confirm the spontaneous breaking of the custodial symmetry in the latter.

\begin{figure}[htb!]
    \centering
    \includegraphics[width=0.49\textwidth]{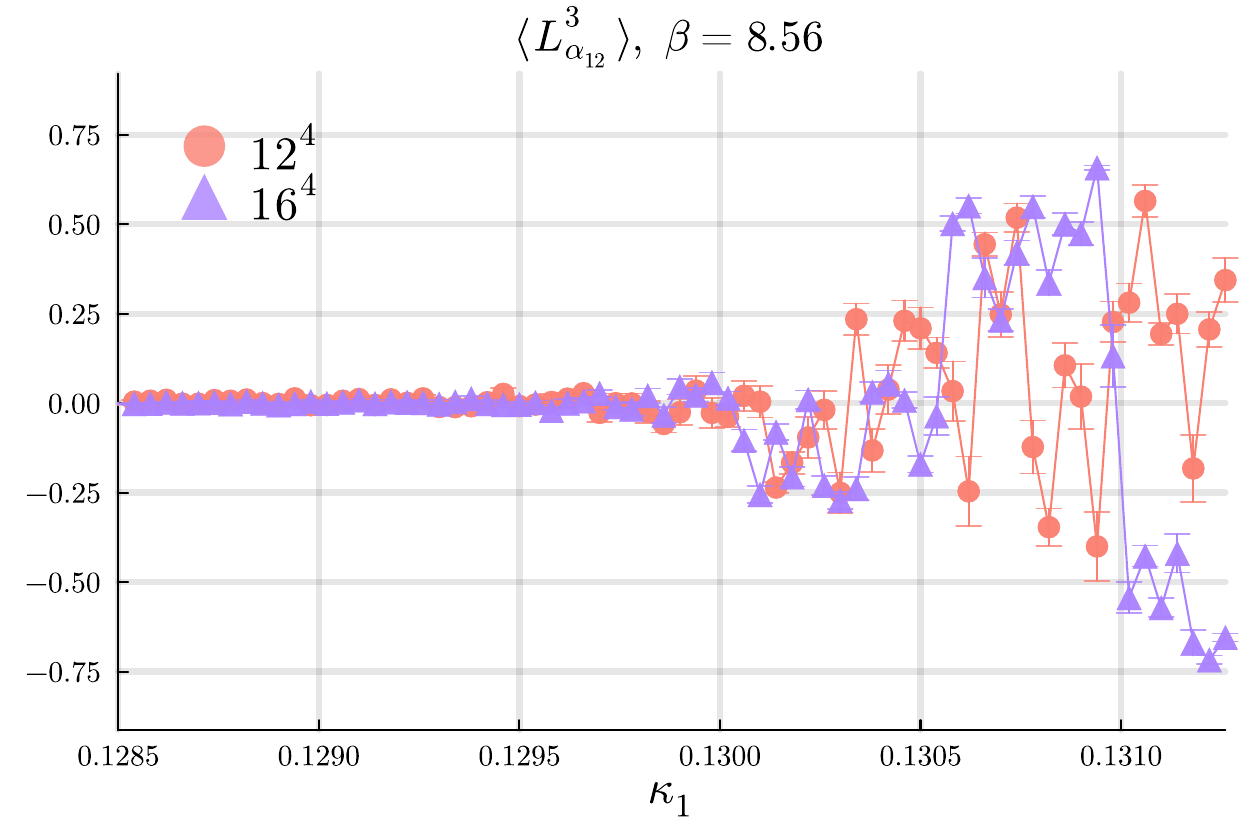}
    \includegraphics[width=0.49\textwidth]{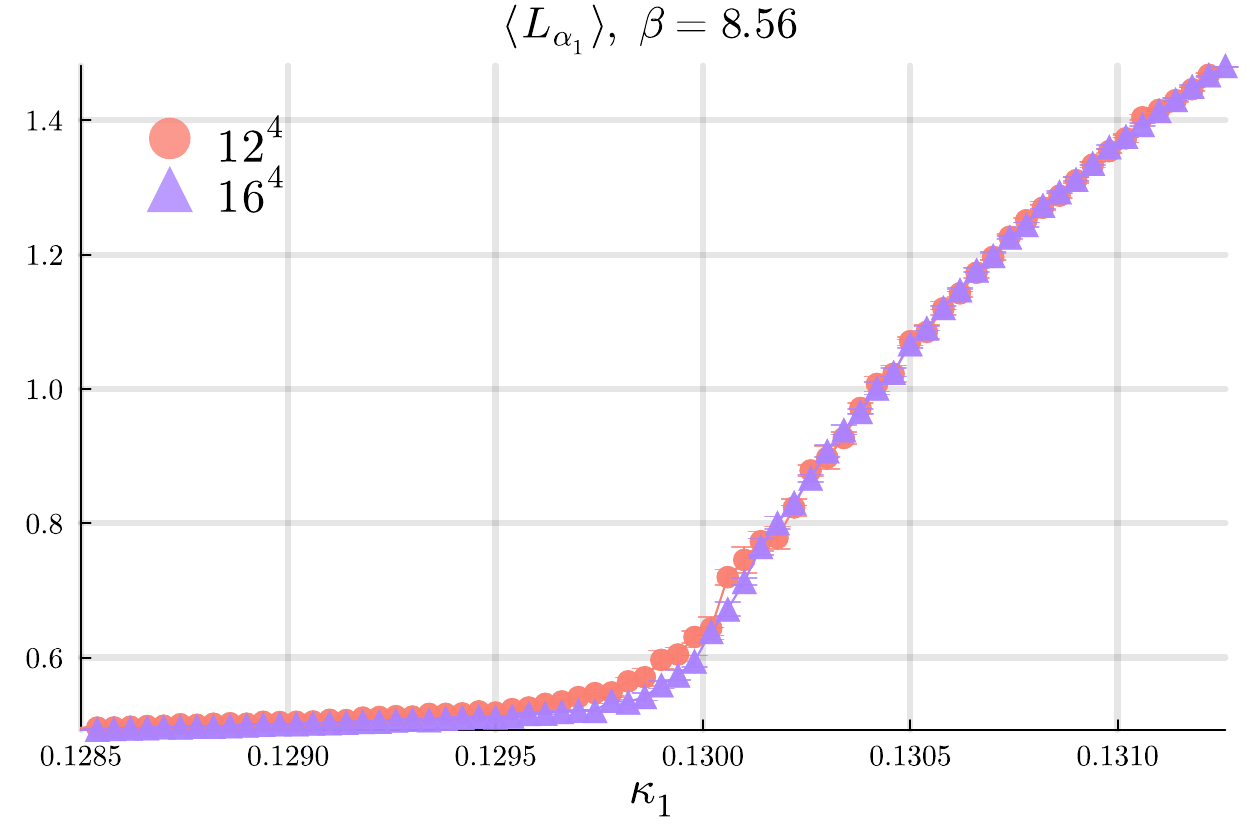}
    \caption{Expectation values for $L_{\alpha_{12}}^{3}$ (left) and
$L_{\alpha_{1}}$ (right) as a function of $\kappa_{1}$ signaling the transition
between $(H_{2})$ and $(H_{12})$. The constant bare couplings are those from the
$\beta=8.56$ point of the LPCP.  Two lattice volumes are shown, $12^{4}$, and
$16^{4}$.}
    \label{fig:k1_scan_global_vevs}
\end{figure}

Similarly to what was found\footnote{A similar approach was
used in this reference to confirm the spontaneous symmetry breaking of
the $SU(2)\times SU(2)$-symmetric 2HDM.} in ref.~\cite{lewis_spontaneous_2010}, 
$\langle L_{\alpha_{12}}^{j} \rangle$ shows large fluctuations with respect to the change of $\kappa_{1}$ when crossing into
$(H_{12})$.  An example of $\langle L_{12}^{3} \rangle$ is demonstrated in the left plot of fig.~\ref{fig:k1_scan_global_vevs}.  This means that care has to be taken when using $\langle L_{\alpha_{12}}^{j}\rangle$ as an order parameter.  As a result, conventional methods
(susceptibility, Binder's cummulant, etc.) may not be straightforwardly applicable for the investigation of this potential phase transition.  In fact, from this plot, it is not possible to draw any reliable conclusion regarding the symmetry breaking pattern.  On the other hand, as exhibited on 
the right-hand side of fig.~\ref{fig:k1_scan_global_vevs}, the plotted quantity, $\langle L_{\alpha_{1}}\rangle \equiv \langle L^{4}_{\alpha_{11}}\rangle$, signals non-trivial changes of the system in the passage from $(H_{2})$ to
$(H_{12})$\footnote{This quantity, $\langle L_{\alpha_{1}}\rangle$, is not an order parameter for
the custodial $SU(2)$ since it is invariant under this symmetry transformation.}.  In order to probe the possible occurrence of the SSB of custodial symmetry, we follow the procedure introduced at the
end of  sec.~\ref{sec:strategy_phase_structure}.  That is, we perform simulations with
an added $SU(2)$-breaking term, $\vareps \mathcal B$ (\cref{eq:ssb_breaking_term}) in the action, and extrapolate $\vareps$ to zero in the thermodynamic limit.

\begin{figure}[htb!]
    \centering
    \includegraphics[width=0.49\textwidth]{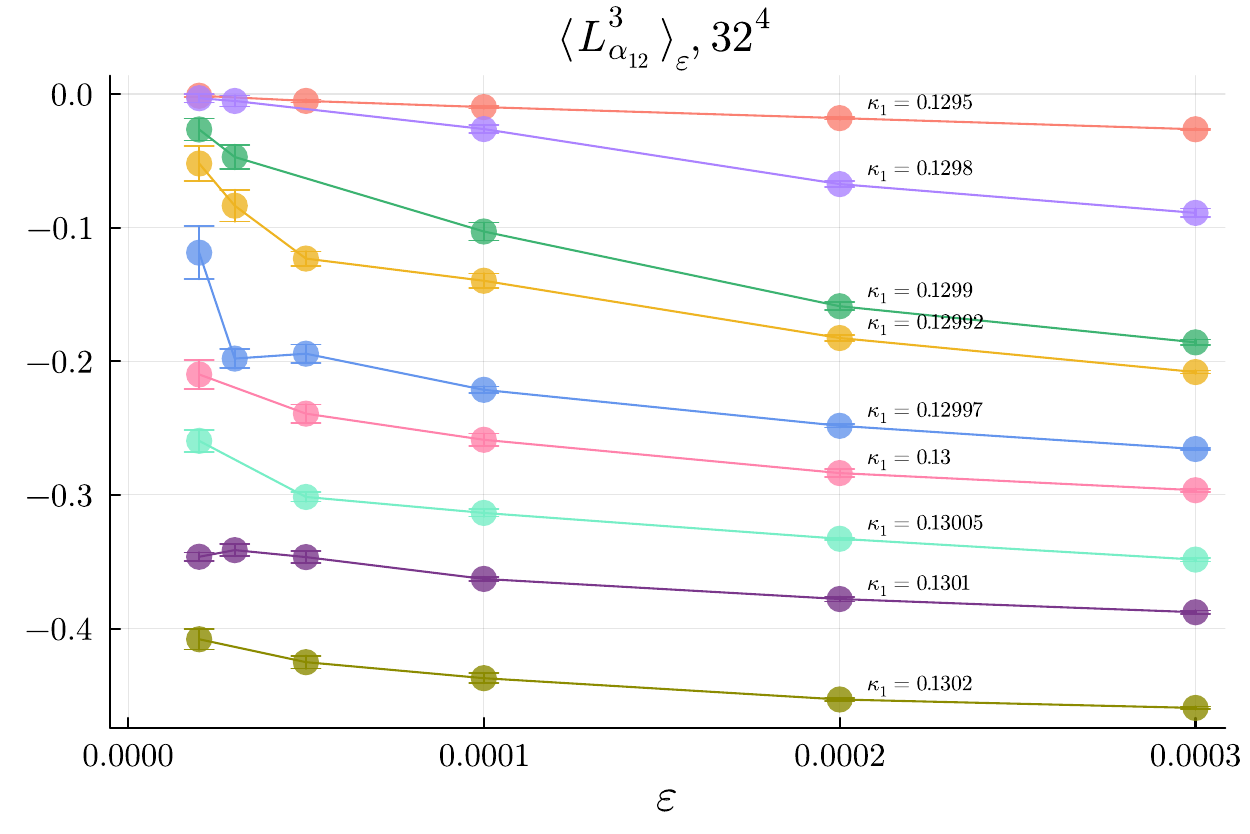}
    \includegraphics[width=0.49\textwidth]{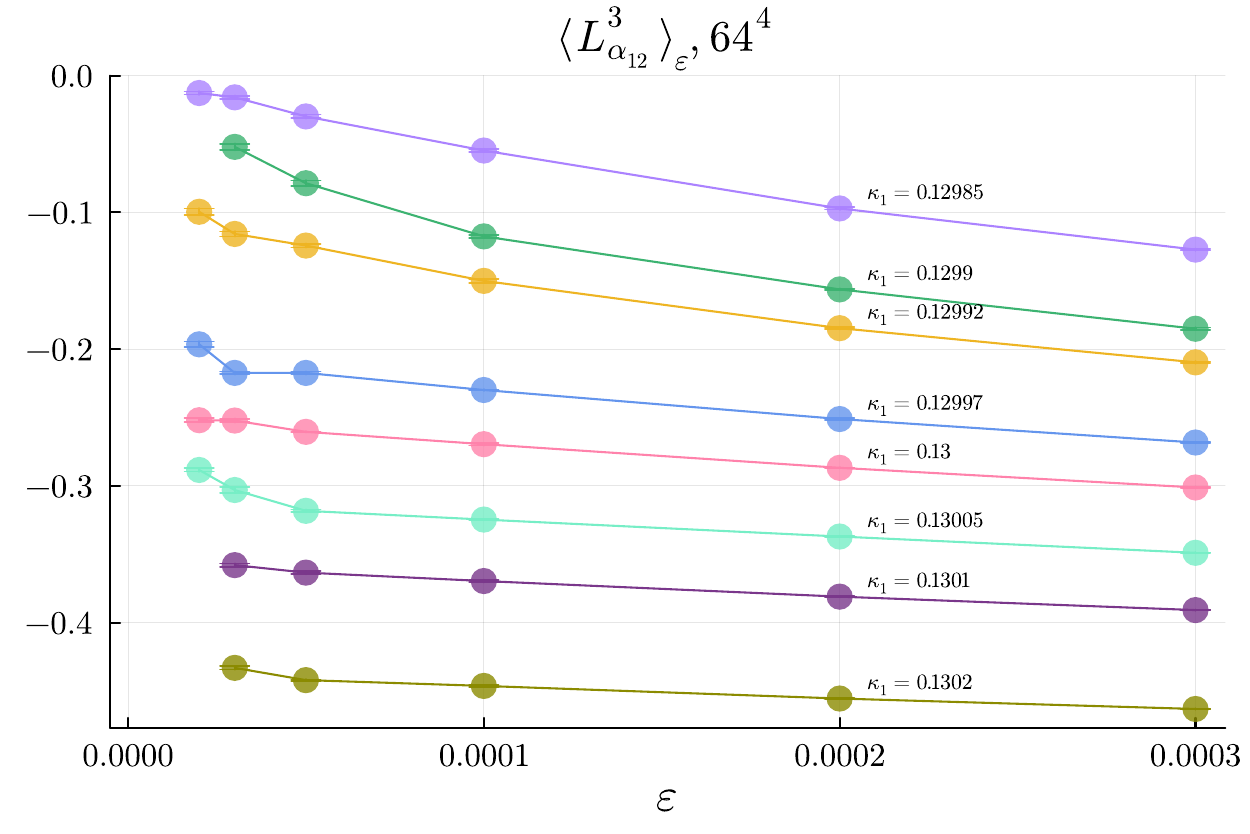}
    \caption{Expectation value of the $SU(2)$ non-invariant quantity
$L_{\alpha_{12}}^{3}$ as a function of the symmetry breaking parameter $\vareps$
for different values of the hopping parameter $\kappa_{1}$ below and above the
transition. Two different volumes are shown.}
    \label{fig:eps_symbreak}
\end{figure}

The strategy is to then perform simulations at differentvalues of $\kappa_{1}$ straddling the boundary between 
sectors $(H_{2})$ and $(H_{12})$ with multiple choices of $\vareps$ and the lattice volume, $L$ ($L=32, 44, 64$).
The study is carried out at relatively high lattice cutoff, with
$\beta=8.56$ in eq.~(\ref{eq:YM action}). For other bare couplings that appear in $S_{\text{2HDM}}$ in eq.~(\ref{eq:lattice action}), we set their values to be those in the $\beta=8.56$ column in 
tab.~\ref{tab:LCP}.  Results of $\langle L_{\alpha_{12}}^{3}\rangle$ from these simulations are shown in
\cref{fig:eps_symbreak} for $L=32$ and $L=64$.  Since at any finite
$\vareps$ the system is forced into a unique ground state chosen by
the inclusion of the $\mathcal B$-term, \cref{eq:ssb_breaking_term},
the expectation value now shows a smooth dependence on
$\kappa_{1}$.  From the plots in fig.~\ref{fig:eps_symbreak}, it is evidenced that $\langle L_{\alpha_{12}}^{3}\rangle$ will extrapolate to a non-zero value in the limit $L \rightarrow \infty$ and $\varepsilon \rightarrow 0$ when $\kappa_{1}$ is above a certain critical value, $\kappa_{1}^{c}$, while it may vanish in the regime $\kappa_{1} < \kappa_{1}^{c}$.  It is noted that at large enough $\kappa_{1}$, finite-volume effects are less significant, where it is also seen that the extrapolation in $\varepsilon$ is less pronounced at large volume.


\begin{figure}[htb!]
    \centering
    \includegraphics[width=0.7\textwidth]{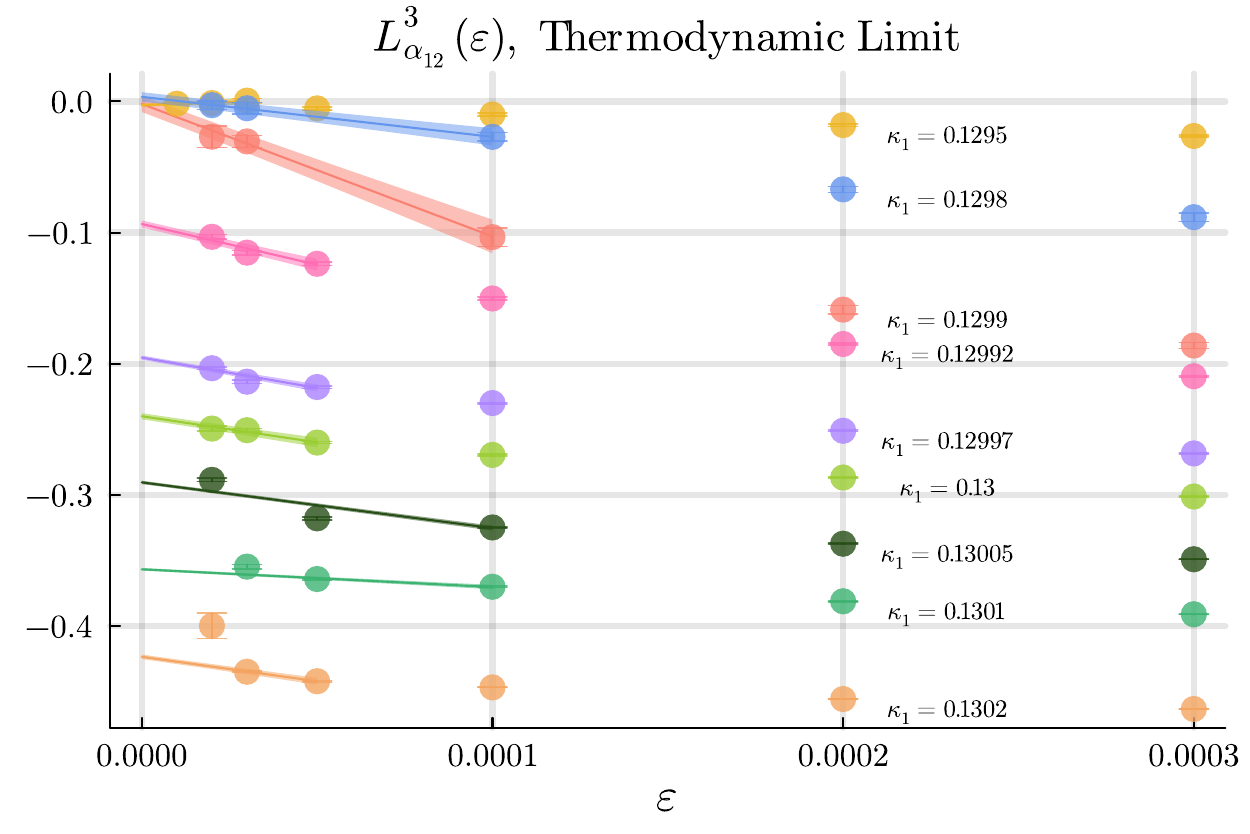}
    \caption{Expectation value of the $SU(2)$ non-invariant quantity
$L_{\alpha_{12}}^{3}$ as a function of the symmetry breaking parameter $\vareps$
in the thermodynamic limit for different values of the hopping parameter,
$\kappa_{1}$.}
    \label{fig:eps_symbreak_infvol}
\end{figure}

The above observations indicate that $\expval{L_{\alpha_{12}}^{3}}$ in the 
$L\rightarrow \infty$ and $\varepsilon \rightarrow 0$ limits can be employed as an order parameter for investigating spontaneous breaking of the global $SU(2)$ symmetry in sector $(H_{12})$.  Following the procedure described in eq.~(\ref{eq:limit_ssb_lattice}), we use our data at $L=32, 44, 64$ to extrapolate $\expval{L_{\alpha_{12}}^{3}}_{\varepsilon}$ to the infinite-volume (thermodynamic) limit first at each pair of $(\kappa_{1},\vareps)$.  Results in this limit are demonstrated in 
\cref{fig:eps_symbreak_infvol}, where we also show the subsequent extrapolation to $\varepsilon \rightarrow 0$ through a linear fit by employing data at the three smallest values of $\varepsilon$ for each $\kappa_{1}$\footnote{The addition of more points in the fit, or the
use of a higher order polynomial does not greatly affect the central
value of the extrapolation for $\kappa_{1}$ close to $\kappa_{1}^{c}$.
On the other hand, the precise shape of the curve for values farther
above $\kappa_{1}^{c}$ is not as relevant for the purpose of this
investigation.  }.


\begin{figure}[htb!]
    \centering
    \includegraphics[width=0.7\textwidth]{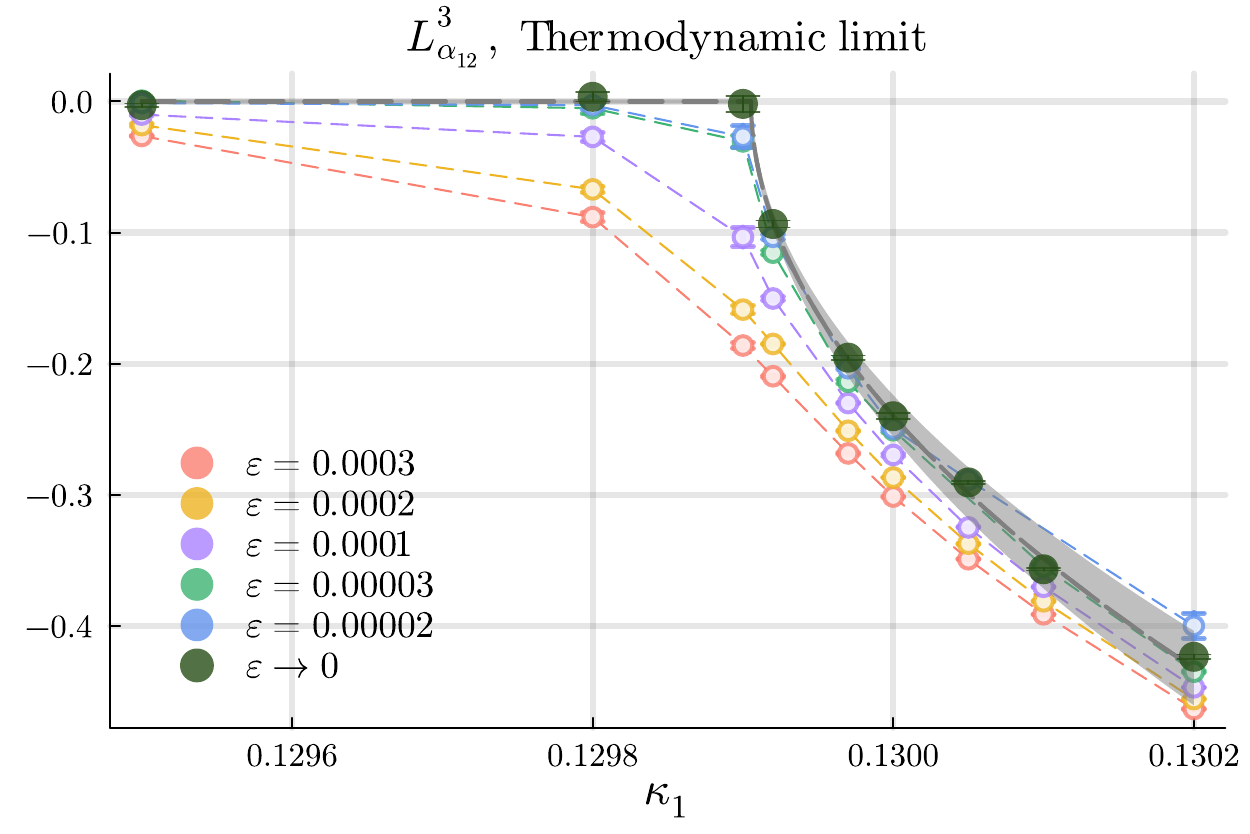}
    \caption{Order parameter $\langle L_{\alpha_{12}}^{3} \rangle$ in the thermodynamic limit
and with the $\vareps\rightarrow0$ extrapolation.}
    \label{fig:ssb_order_param_k1dep}
\end{figure}

Figure~\ref{fig:ssb_order_param_k1dep} exhibits
$\langle L^{3}_{\alpha_{12}}\rangle$ as a function of $\kappa_{1}$, at
$\varepsilon = 0$ and in the thermodynamic limit.  Using data points
displayed in this plot, we can estimate the value of $\kappa_{1}^{c}$.
In order to do this, we parametrize the $\kappa_{1}$-dependence in the
broken phase with a functional form inspired by continuous
transitions, that is,
$\expval{L^{3}_{\alpha_{12}}} = A(\kappa_{1}-\kappa_{1}^{c})^{\nu}$.
The result of the fit to this functional form is also shown in
\cref{fig:ssb_order_param_k1dep}, with the parameters
$\kappa_{1}^{c}=   0.1299053(13)$ and $\nu=       0.5184(80)$.  It is
important to remark that although we have employed a functional form
from continuous phase transitions to parametrize the order parameter
in \cref{fig:ssb_order_param_k1dep}, the type of phase transition is
not determined.\footnote{The use of this functional form serves only
to obtain a rough estimate of $\kappa_1^c$, but a precise estimate of
this value is not relevant for the study. Similarly, the use of
smaller $\varepsilon$ would not change the qualitative conclusions but
serve only to improve the resolution of the transition.} In fact, as
will be seen in  \cref{sec:results_spectrum}, the masses of the states
in lattice units in the theory generally do not vanish in the infinite
volume limit across the $(H_{2})\rightarrow(H_{12})$ transition,
indicating that the correlation length does not diverge, thus
excluding the possibility of a second-order phase transition.  While
the exploratory results in ref.~\cite{Catumba:2025obz} seem to
indicate a first-order transition, the assessment of the type of
transition is beyond the scope of this work.

\subsection{Tuning the standard model line of constant physics and setting the scale}
\label{sec:results_tuning}

In the previous subsection, the absence of the custodial symmetry in the spectrum in sector
$(H_{12})$ was established. It is also known that the Higgs/FMS mechanism is inactive in
$(H_{0})$.  This leaves $(H_{1})$ and $(H_{2})$ to be the only relevant regimes of the parameter space for model building, since these are the sectors where the SM can be embedded.  In this work we consider the LPCP (see sec.~\ref{sec:strategy_phase_structure_constant_physics}) in sector $(H_{2})$, with $\kappa_{2}>\kappa_{2}^{c}$ and
$\kappa_{1}<\kappa_{1}^{c}$.  This means that the SM states couple to
the $\Phi_{2}$ interpolators as described in
sec.~\ref{sec:spectrum_interpolators}. 
 Other than the SM states, in this
sector we find four non-SM scalar states with masses $m_{H}$ and
$m_{A}=m_{H^{\pm}}$.  From eq.~(\ref{eq:interpolators_generic}), the correspondence between the interpolators and the
states is (the superscript index $j$ runs from 1 to 3)
\begin{align*}
  \begin{aligned}
    &S_{22}\, ,\\
    &W_{22}^{j}\, ,\\
    &S_{12}^{4}\, ,~W_{12}^{4}\, ,\\
    &S_{12}^{j}\, ,~W_{12}^{j}\, ,
  \end{aligned}
  &&
  \begin{aligned}
    &\textrm{SM Higgs}~(m_{h})\, ,\\
    &W\textrm{-bosons}~(m_{W})\, ,\\
    &\textrm{BSM scalar}~(m_{H})\, ,\\
    &\textrm{BSM scalars}~(m_{A}=m_{H^{\pm}})\, .
  \end{aligned}
\end{align*}

Our strategy to build the LPCP is to sequentially
increase the bare coupling $\beta$ in eq.~(\ref{eq:YM action}) towards finer lattice spacing. At each $\beta$ value, we scan the
$\{\kappa_{2},\eta_{2}\}$-space to find parameters such that the SM conditions,
\begin{align}
  &R = \left(\frac{m_{h}}{m_{W}}\right)_{\textrm{latt}} = \left(\frac{m_{h}}{m_{W}}\right)_{\textrm{phys}} = 1.5\label{eq:R=1.5}\, ,&&\\
  &S = \left(\frac{m_{W}}{\mu_{0}}\right)_{\textrm{latt}} = \left(\frac{m_{W}}{\mu_{0}}\right)_{\textrm{phys}} = 1.0\, ,  \mbox{ }\mathrm{with}\mbox{ }g_{\textrm{GF}}^{2}(\mu_{0})_{}=0.5\, ,~{\rm and} \mbox{ }\mu_{0}=\frac{1}{\sqrt{8t_{0}}}\, ,\label{eq:S=1}
\end{align}
are satisfied.\footnote{It is important to remark that the second
physical condition is valid in the $\overline{\rm{MS}}$-scheme, and
thus our condition is only valid at leading order. However, given the
targeted precision and  the small values of the couplings, we
expect this condition to represent SM physics.  } Additionally, in this work the
lattice spacing in natural units is obtained by
\begin{align}
  \label{eq:scale_setting}
  a = \frac{(am_{W})^{\rm latt}}{m_{W}^{\rm phys}}\, ,
\end{align}
where $m_{W}^{\textrm{phys}}=80.377(12)~\si{GeV}$ is the physical
value of the $W$ boson mass \cite{Workman:2022ynf}, and $(am_{W})^{\rm latt}$ the
corresponding lattice measurement.

While we adjust $\kappa_{2}$ and $\eta_{2}$ to tune for the above SM conditions, the remaining couplings ($\kappa_{1},\eta_{1}, \eta_{3}, \eta_{4}=\eta_{5}$) in the action in eq.~(\ref{eq:lattice action}) are fixed to constant values, such that the
BSM scalar masses, $m_{H}$ and $m_{A}=m_{H^{\pm}}$, are larger than the SM particle masses,
but still well below the lattice cutoff.  For this purpose,
$\kappa_{1}$ and $\eta_{1}$ are chosen such that the system is well within the $(H_{2})$ sector.  That is, the value of $\kappa_{1}$ is much smaller than
$\kappa_{1}^{c}$ ($\kappa_{1} = 0.1245$ and $\eta_{1}=0.003$).  Finally, the couplings $\eta_{3},\eta_{4}=\eta_{5}$ are
kept small ($\eta_{3}=0.0002$, $\eta_{4}=\eta_{5}=0.0001$).   This guarantees the LPCP
to be mostly insensitive to the $\Phi_{1}$ scalar sector (within the available
precision).  Notice that while we fix the values of all the BSM bare couplings
for the investigation of LPCP in this subsection, later on we will perform scans in $\kappa_{1}$ to explore some limits of the BSM spectrum (see sec.~\ref{sec:results_spectrum} below).

For each $\beta$ value, around ten replicas of $\sim10K$ HMC
trajectories are generated in our simulations.  Measurements of the interpolators are
performed at each HMC trajectory, while the computation of the gradient flow
action density is carried out less frequently (the frequency is adjusted
for each $\beta$ value).  For the finest lattice the resulting
integrated autocorrelation time for the Higgs mass is
$\tau_{i,m_{h}}\approx 3$ molecular dynamics units (MDUs).

\begin{table}
  \centering
  \begin{tabular}{cccccc}
    \toprule
 $\beta$                &        8.2 &        8.3 &        8.4 &       8.56 &       8.64 \\
 $\kappa_2$             &    0.13175 &    0.13104 &     0.1306 &     0.1301 &   0.129985 \\
 $\eta_2$               &    0.00338 &      0.003 &    0.00285 &    0.00275 &   0.002737 \\
    \midrule
 $R$                    &  1.509(94) &  1.527(80) &  1.494(65) &  1.504(39) &  1.462(53) \\
 $S$                    & 1.0055(98) & 0.9956(65) &  0.994(14) &  0.994(20) & 0.9958(94) \\
 $am_{h}$               &  0.402(28) &  0.363(21) &  0.305(16) & 0.2192(84) & 0.1863(75) \\
 $am_{W}$               & 0.2666(26) & 0.2377(15) & 0.2041(29) & 0.1458(29) & 0.1275(11) \\
 $t_0/a^2$              & 1.7781(58) & 2.1934(74) &  2.966(13) &  5.810(45) &  7.626(43) \\
 $1/a~(\si{GeV})$ &  301.5(29) &  338.2(21) &  393.8(57) &    551(11) &  630.4(56) \\
 $m_{W}L$                   &  8.485(14) &  7.639(13) &  6.569(15) &  4.694(18) &  6.145(17) \\

    \bottomrule
  \end{tabular}
  \caption{Bare couplings $\beta,\kappa_{2},\eta_{2}$ of the LPCP together with
the corresponding physical conditions $R,S$, the SM masses $am_{h},am_{W}$ in
lattice units, the gradient flow scale $t_{0}/a^{2}$ and the estimated cutoff
energy $\Lambda_{c} = 1/a$. The remaining couplings were fixed to the values:
$\kappa_1=0.1245$; $\eta_1=0.003$; $\eta_{3}= 0.0002$; $\eta_{4}=\eta_{5}=0.0001$.
The simulations are performed on lattices with sizes $L=28,28,32,32,48$ for the five $\beta$ values.}
  \label{tab:LCP}
\end{table}

The results for renormalization trajectory in the 3-dimensional parameter space spanned by 
$(\beta,\kappa_{2},\eta_{2})$ are dispayed in \cref{tab:LCP} .  Five
$\beta$ values are considered, with the cutoff, $1/a$ computed from
\cref{eq:scale_setting},  ranging from $300~\si{GeV}$ to
$630~\si{GeV}$.  The simulations are performed with large enough
volumes, $L=32, 32, 32, 32, 48$ for the five $\beta$ values, such that finite volume effects are
mitigated. In fact, the value $m_{W}L$ is always kept above 4 (for
this set of couplings, $m_{W}$ is the lowest mass in the spectrum).

\begin{figure}[htb!]
  \centering
		\includegraphics[width=0.49\textwidth]{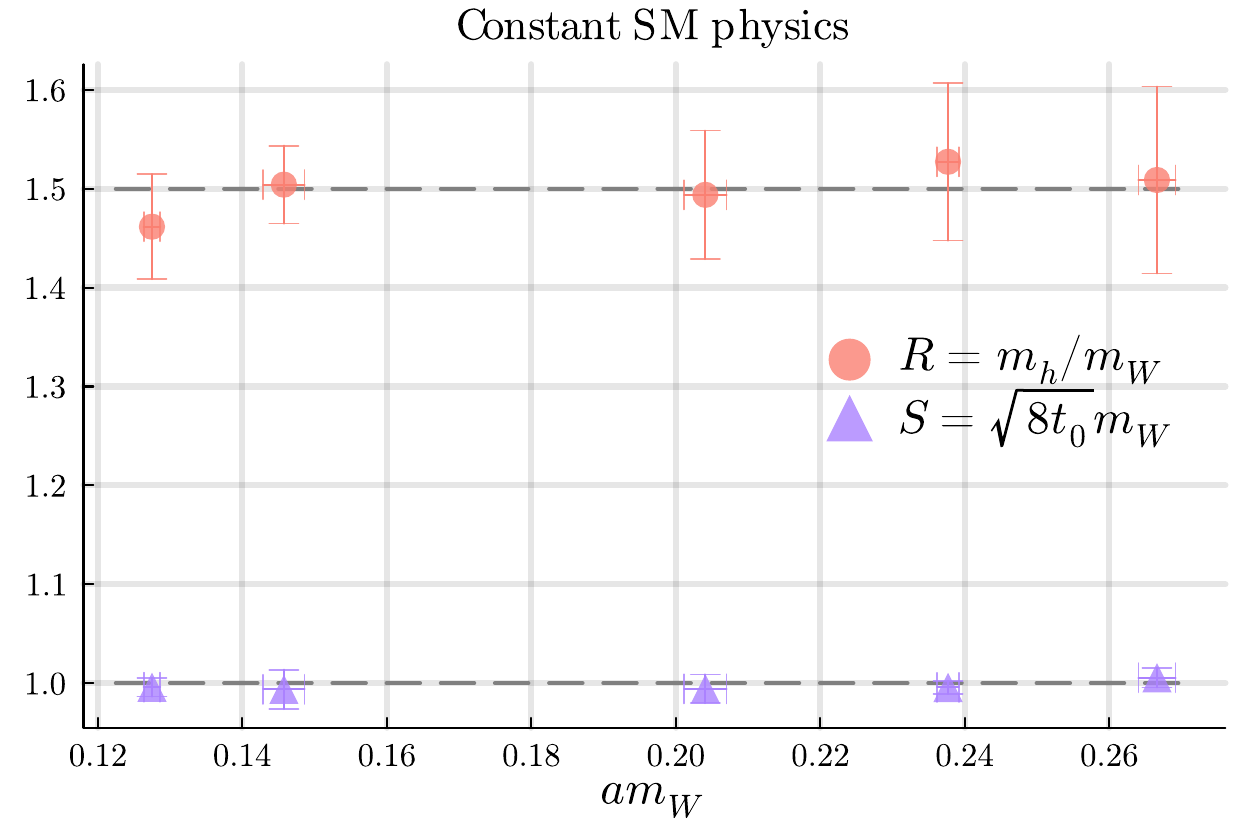}
    \includegraphics[width=0.49\textwidth]{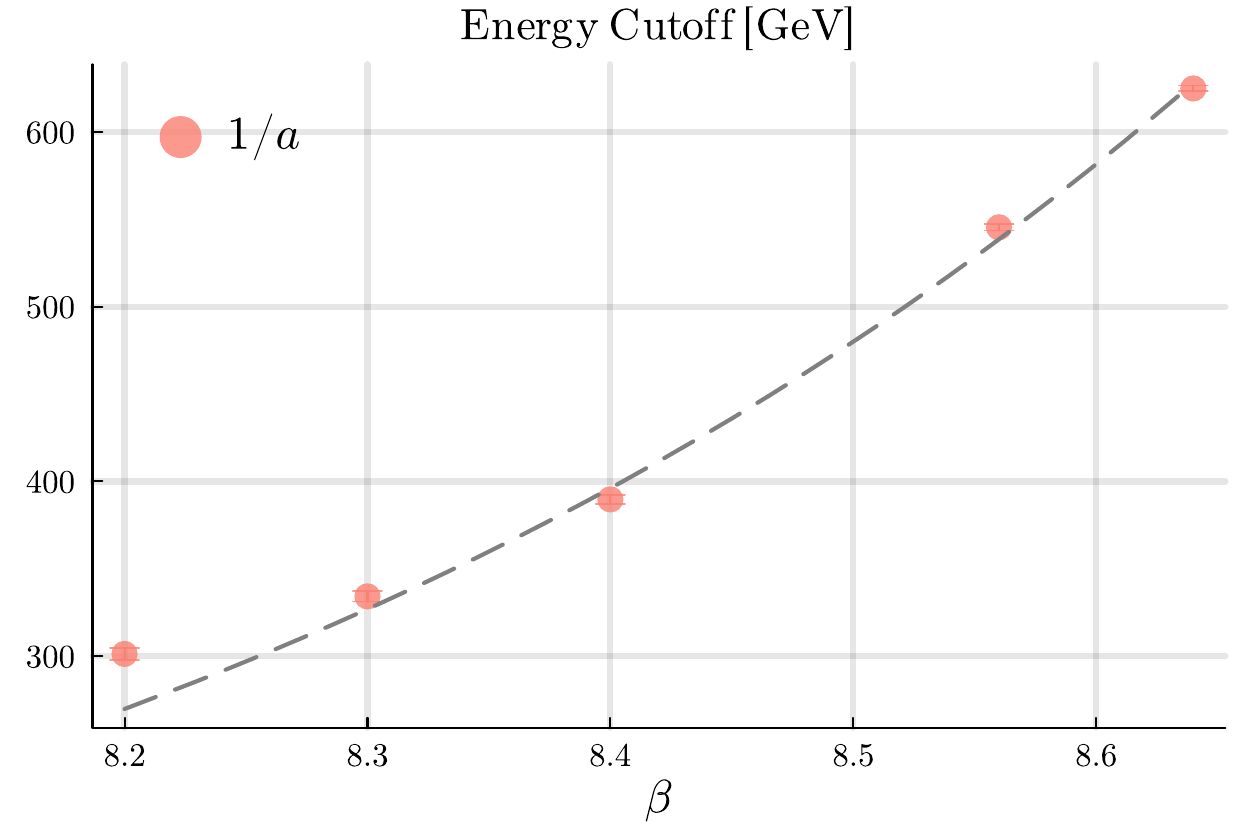}
		\includegraphics[width=0.49\textwidth]{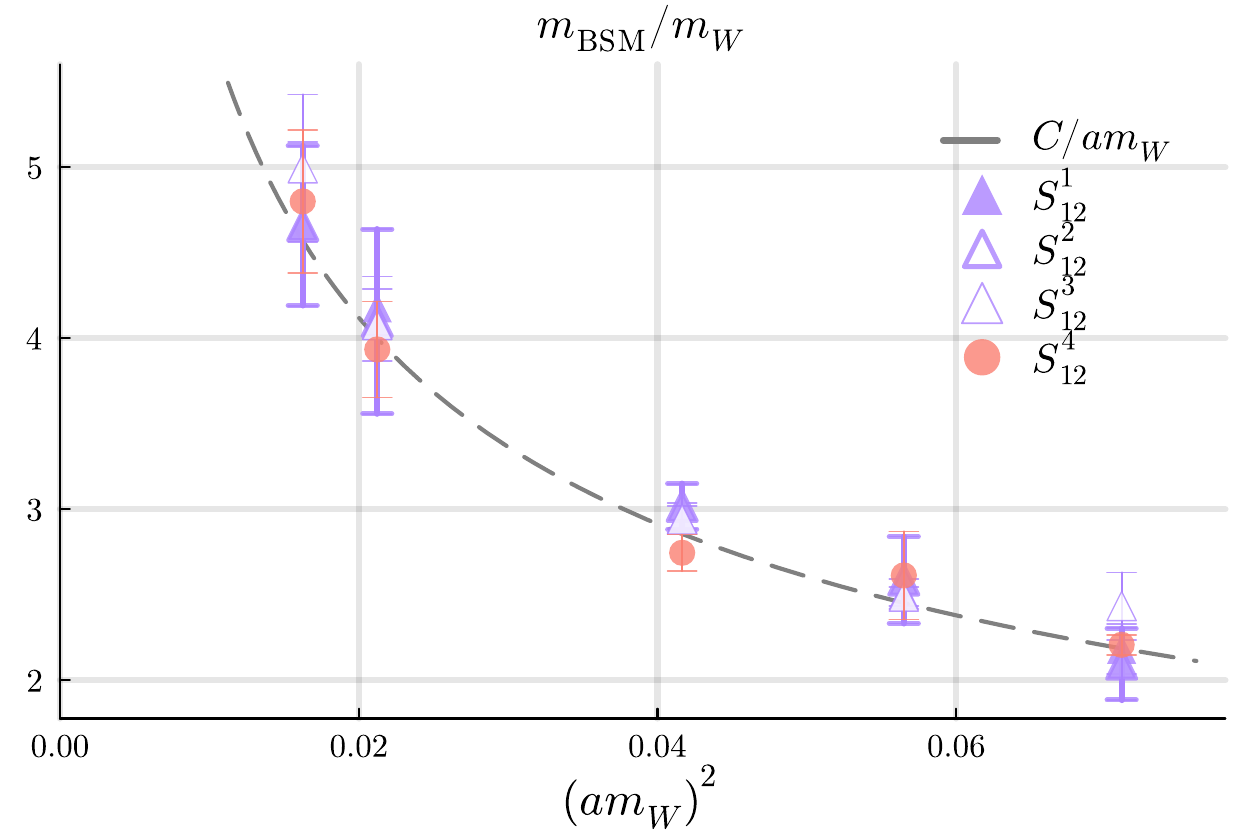}

\caption{Data from \cref{tab:LCP}: physical conditions $R$ and $S$ for
the selected points in the line of constant SM physics as a function
of $am_{W}$ (left).  For decreasing $am_{W}$, each points has a
corresponding increasing $\beta$ value.  The lattice cutoff,
$1/a$, estimated from $a=\hat m_{W}/m_{W}^{\textrm{phys}}$ is
shown on the right as a function of $\beta$. The bottom plot shows the
ratio of the BSM states $m_{H^{\pm}}=m_{A}$ (triangles), and $m_{H}$
(circle) to the $W$ boson mass for the points in the LPCP. The fit to the
functional form $C/am_{W}$ (dashed line), with $C$ a fitting constant,
is only meant to indicate that, within errors, the increase in the
ratio $m_{\rm BSM}/m_{W}$ seems to come mostly from the change in $am_{W}$
towards the continuum. 
}
\label{fig:lcp}
\end{figure}

Both the physical conditions for $R$ and $S$ in eqs.~(\ref{eq:R=1.5}) and (\ref{eq:S=1}), and the lattice cutoff, $1/a$, are
shown in the upper row of \cref{fig:lcp} as a function of $am_{W}$ and $\beta$,
respectively.  The difference in precision of condition $R$ and that of condition $S$ is clear in the left plot.  This is because the computation of $R$ involves the measurements of scalar masses.
Nonetheless, both conditions are satisfied within errors for all
points.  On the right-hand side the cutoff seems to follow an
exponential dependence in $\beta$, as expected in the scaling region.

Since $\kappa_{1}$ is kept fixed in our construction of LPCP, the BSM masses in lattice units are expected to be 
roughly constant on the five ensembles listed in \cref{tab:LCP}.  In fact, it is found that $am_{H} \approx am_{A} \sim 0.6$ with large
uncertainties.  Furthermore, in the bottom plot of \cref{fig:lcp}, the mass ratio between various 
BSM scalar states and the $W$ boson is shown as a function of $am_{W}$. 
This plot shows that this ratio grows when the lattice spacing becomes finer, since the SM physics is embedded ($R=1.5$ and $S=1$) in these five data points.  
This
simply indicates that these points do not define a `full renormalization
trajectory' of the 2HDM.  In order to obtain the complete line of constant physics for this model, it would be necessary to perform further tuning of the
remaining BSM couplings, such that at each $\beta$ value, $m_{\rm BSM}/m_{W}$ is kept fixed.


The results listed in tab.~\ref{tab:LCP} show that the quartic
coupling $\eta_{2}$
decreases as the cutoff scale increases (or the lattice spacing becomes finer).   This seems to be the opposite of what one expects from triviality which predicts that bare quartic couplings grow when the lattice spacing becomes finer along a line of constant physics, and eventually diverge before the continuum limit is reached~\cite{Montvay_Münster_1994}. 
This could, at the first glance, suggest
the existence of a non-perturbative, non-trivial fixed point with
renormalized $\eta_{2} \neq 0$ in the continuum. However, it is important to
remark that triviality is usually framed within mass-independent
renormalization schemes for examining the properties of the quartic couplings.  Contrarily, in this work we set up a LPCP non-perturbatively, using a renormalization scheme
that is inherently massive.  
Such a scheme may lead to a different
running behaviour of the quartic coupling compared to that in a mass-independent scheme. 
A clear example will be
seen in the following subsection where we study the running of the
gauge coupling.  We leave a dedicated, non-perturbative
investigation of triviality in gauge-Higgs theories for our future work.


%

%
\subsection{Running gauge coupling}
\label{sec:results_gauge_coupling}

\begin{figure}[htb!]
  \centering
  \includegraphics[width=0.7\textwidth]{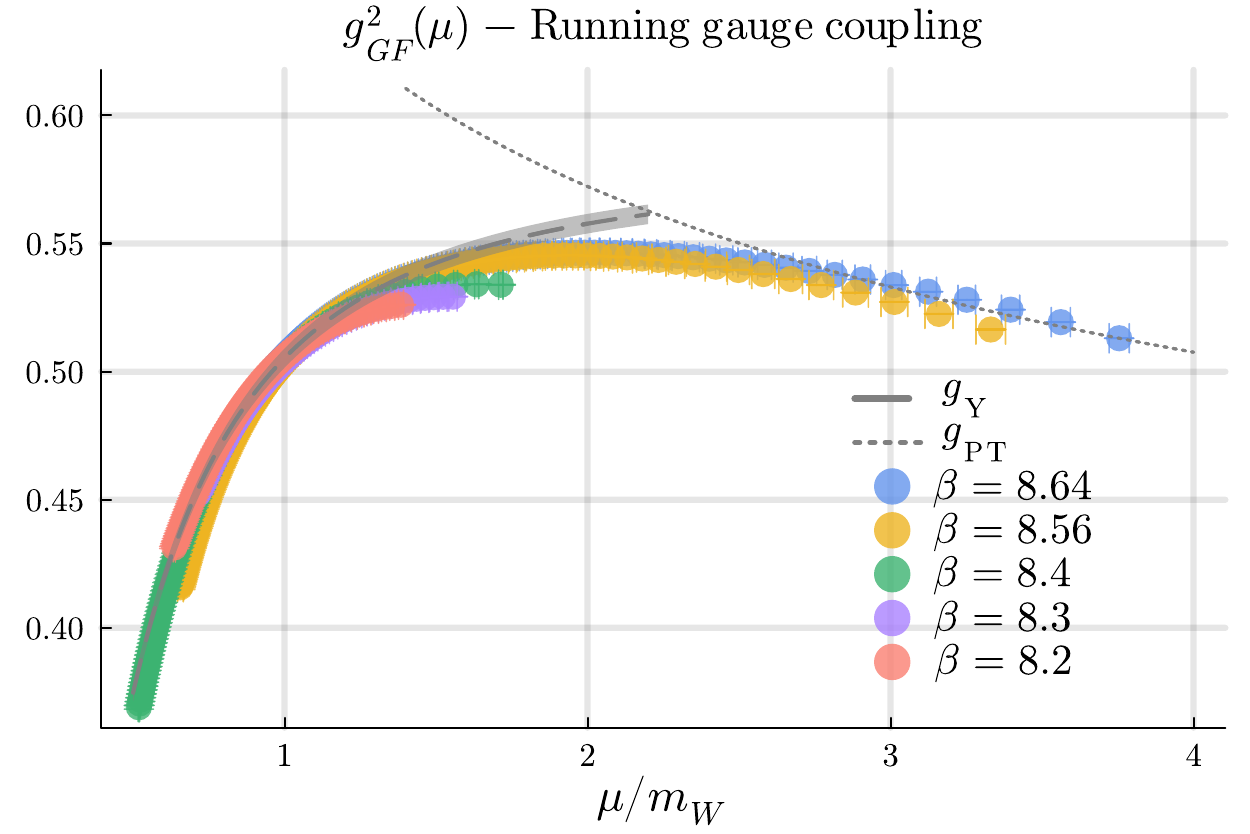}
  \caption{Running behaviour of the gauge coupling, $g_{GF}^{2}(\mu)$, 
from lattice simulations along the LPCP in \cref{tab:LCP}. Only results with
$\mu a < 0.5$ are shown. The one-loop perturbative prediction, $g_{\rm PT}^{2}$, is matched to 
$g_{GF}^{2}(\mu)$ computed at $\beta = 8.64$ at $\mu/m_{W} \approx 3.5$.
 The gauge coupling from the Yukawa
potential is fitted in the infrared region, with the estimated screening mass
$m_{\textrm{screen}}= 0.6243(28)m_{W}$.}
  \label{fig:running_coupling}
\end{figure}

In this work, the running of the renormalized gauge coupling is studied in the gradient-flow scheme defined through the
action density, as described in \cref{eq:GF_coupling}.  Results of $g_{GF}^{2}(\mu=1/\sqrt{8t};\beta)$ for all five  $\beta$ values listed in \cref{tab:LCP}
 are exhibited in
\cref{fig:running_coupling} as a function of the energy scale, $\mu$, relative
to $m_{W}$.  
For each lattice spacing, we only show results at $\mu a < 0.5$, corresponding to $t/a^{2}>0.5$, since lattice artefacts can be large for those at smaller flow time. 
Figure~\ref{fig:running_coupling} demonstrates that $g_{GF}^{2}(\mu=1/\sqrt{8t};\beta)$ obtained at different $\beta$ values are compatible in a large range of $\mu/m_{W}$, with significant lattice artifacts
being observed only near $t/a^{2}=0.5$. 
Notice that in the tuning for the LPCP, the uncertainty in the `mass condition', $R$, is larger than that in $S$, as shown in~\cref{fig:lcp}. However, the results in~\cref{fig:running_coupling} seem to indicate that effects of the potential mistuning of the masses are negligible in this quantity.

We can compare our numerical study with the prediction of perturbation theory, where the latter is expected to be valid, that is, at large $\mu$.
In~\cref{fig:running_coupling} the one-loop perturbative result, $g_{\rm PT}^{2}(\mu)$, is also
shown.  The value of $g_{\rm PT}^{2}(\mu)$ is first matched to that of $g_{GF}^{2}(\mu)$ computed on the $\beta = 8.64$ ensemble in~\cref{tab:LCP}, at a renormalization scale well above $m_{W}$.  We choose this matching scale to be $\mu\approx 3.5 m_{W}$.  The running behaviour of $g_{\rm PT}^{2}(\mu)$ is then obtained  
from the one-loop $\beta$-function,
\begin{equation}
  \label{eq:betafunction}
    \beta_{SU(N)+\rm Scalars}= \mu \dv{g}{\mu} = -\frac{b_0g^3}{16\pi^2}+\order{g^5}\, ,~~b_0=\frac{22N-n_s}{6} \, ,
\end{equation}
where $n_{s}=2$ is the number of scalar-field doublets, and $N=2$ for $SU(2)$ gauge theory. 
 Notice
that the lattice data and the perturbative curve overlap only in a 
limited range since the massless perturbative scheme is only valid at
very large $\mu$.  To further extend this overlapping regime, one has to perform lattice simulations at finer lattices, which is beyond the scope of this work.

The behaviour of the running gauge coupling presented in 
\cref{fig:running_coupling} can be
understood as follows. When the renormalization scale is well above $m_{W}$, the gauge
bosons are effectively massless.  Consequently, the
gauge coupling decreases as we increase the scale.  The theory is then QCD-like, with the onset of
asymptotic freedom at large $\mu$.  When the energy scale approaches $m_{W}$ from
above, the masses of the gauge bosons become relevant and the coupling stops
increasing.  
The screening effect on the gauge force due to the massive $W$
boson is evident at intermediate values of $\mu$, where we observe significant difference between $g_{\textrm{PT}}^{2}$, obtained in a mass-independent scheme, and 
$g_{GF}^{2}(\mu)$, computed in a mass-dependent scheme.  
Notice that this effect is already
relevant around $\mu/m_{W}\approx 2$, where $g_{GF}^{2}(\mu)$ exhibits very mild dependence on $\mu$.

In the low-energy regime where $\mu \ll m_{W}$, the gauge
boson is expected to decouple completely.  Here the model effectively becomes a scalar field theory, and the coupling decreases with $\mu$, as can be seen in~\cref{fig:running_coupling}.  
In this region, the gauge interaction
can be characterized by an Yukawa-like potential
\cite{langguth_monte_1986,fodor_simulating_1994}.  
In fact, the first lattice
studies of the gauge coupling in $SU(2)$-Higgs models were carried out using the static potential
of a pair of external $SU(2)$ charged doublets.  For large enough separations of
the charges, the potential is expected to be Yukawa-like due to the exchange of
a massive boson.  In the continuum the potential is given by the integral
\begin{equation}
    V_{\textrm{Y}}(r)=\int \frac{d^3k}{(2\pi)^2} \frac{e^{ik\cdot r}}{k^2+m^2} = \frac{1}{4\pi r} e^{-mr} \, ,
\end{equation}
involving the propagator of the massive boson.
We use this to describe the running gauge coupling at $\mu \ll m_{W}$,
\begin{equation}
  g_{\textrm{Y}}^{2}(\mu) = \eval{r^{2}\dv{V_{\textrm{Y}}}{r}(r)}_{\mu = 1/r} \, ,
\end{equation}
whose functional form can be employed to fit the lattice data.  
The result of
the fit for the finest lattice is shown for the low-energy region in~\cref{fig:running_coupling}, with the corresponding Debye
\textit{screening-mass} estimated to be
$m_{\textrm{screen}}= 0.6243(28)m_{W}\approx 50~\si{GeV}$.    The
relevant point here is the qualitative agreement between the running behaviours of $g_{\textrm{Y}}^{2}(\mu)$ and $g_{GF}^{2}(\mu)$ 
for scales up to $m_{W}$. On the
other hand,  although it may be related to the W mass,
\cite{fodor_simulating_1994,Csikor:1999ft}, the screening-mass
obtained from the fit is not a physical quantity.\footnote{This screening mass clearly depends on the choice of renormalization scheme.}

\subsection{Spectrum of scalar states beyond the standard model}
\label{sec:results_spectrum}

The tuning for the LPCP in tab.~\ref{tab:LCP} is performed by adjusting the set of three bare couplings, $\{\beta,\kappa_{2},\eta_{2}\}$, spanning the parameter space for the SM Higgs sector in the presence of $SU(2)$ gauge fields.  In this procedure, all the other couplings (the BSM couplings) were held fixed, as explained in sec.~\ref{sec:results_tuning}.  The values of these BSM couplings actually have profound implications for the phenomenology of the custodial 2HDM under investigation.  This subsection describes our exploration of 
physically realizable scenarios by
varying them, while keeping to the LPCP. In particular, we are interested in the objective of establishing
non-perturbative bounds on the BSM spectrum of this model. This
would, in principle, require a scan in the space of all the remaining parameters, $\{\kappa_{1}, \eta_{1}, \eta_{3},\eta_{4}\}$, while maintaining the conditions of eqs.~(\ref{eq:R=1.5}) and (\ref{eq:S=1}). 
 The computational cost of this scan and the
requirement of retuning for the LPCP make it a challenging task.  Consequently, for our first step of investigating the custodial 2HDM reported in this article, the BSM quartic couplings, $\eta_{1,3,4}$, are fixed to be weak, with the values used in our tuning for the LPCP (sec.~\ref{sec:results_tuning}).  We then explore the effect of
changing the hopping parameter of the `unbroken' scalar, $\kappa_{1}$,
within the sector $(H_{2})$.  In principle, varying
$\kappa_{1}$ should require the retuning of
$\{\beta,\kappa_{2},\eta_{2}\}$. However, since $\eta_{1,3,4}$ are kept small, SM physics is expected to be almost insensitive to the change of $\kappa_{1}$.

Before presenting details of our numerical results on the masses of the BSM scalar states, we notice that some qualitative features of the spectrum can be obtained using perturbative results in \cref{eq:tl_mass_Hpm,eq:tl_mass_a,eq:tl_mass_matrix}, since our calculations are performed in the regime of weak quartic couplings.  We expect that the states $A$ and $H^{\pm}$ are degenerate.  The difference between their mass and that of $H$ is dominated by the size of $\eta_{4}$.  All these masses should decrease with increasing $\kappa_{1}$ in sector $(H_{2})$.  Furthermore, when $\kappa_{1} > \kappa_{1}^{c}$, the model is in sector $(H_{12})$ and the $SU(2)$ custodial symmetry is spontaneously broken (sec.~\ref{sec:results_spontaneous_sym_breaking}), where the three degenerate states become Goldstone modes.  The above considerations indicate that we can search for a qualitative lower bound of the BSM masses in sector $(H_{2})$ with $\kappa_{1}$ close to $\kappa_{1}^{c}$.  A more detailed, precise search for this lower bound should be carried out by also varying $\eta_{4}$, which we leave for future work.

\begin{table}[htb!]
\centering
  \begin{tabular}{cccccc}
      \toprule
 $\beta$          &         8.2 &         8.3 &         8.4 &        8.56 &        8.64 \\
      \midrule
 $\kappa_{1}^{c}$ & 0.13089(31) & 0.13018(42) & 0.13021(65) & 0.12996(59) & 0.12965(61) \\
      \bottomrule
  \end{tabular}
  \caption{Transition points, $\kappa_{1}^{c}$, between sector $(H_{2})$ and
$(H_{12})$ for the points in \cref{tab:LCP}, obtained from hysteresis cycles in
$L_{\alpha_{1}}$.  The errors are estimated from the width of the hysteresis curve}
  \label{tab:k1_crit}
\end{table}

In order to perform scans in the $(H_{2})$ sector only, we first
determine $\kappa_{1}^{c}$ for each $\beta$
value in tab.~\ref{tab:LCP}, through examining thermal cycles.  This is carried out by changing slightly $\kappa_{1}$ after performing a small number of HMC updates, and then continuing the simulations.  Implementing a sequence of such changes, one can increase and then decrease $\kappa_{1}$ across
the boundary between sectors 
$(H_{2})$ and $(H_{12})$.
This thermal cycle leads to
hysteresis effects on the global observables such as $L_{\alpha_{1}}$,
from which the location of the phase transition can be estimated.  The
results for $\kappa_{1}^{c}$  are summarized in~\cref{tab:k1_crit}. 
 Notice that for $\beta=8.56$ this rough estimate agrees with the
computation from \cref{sec:results_spontaneous_sym_breaking}.

\begin{figure}[htb!]
\centering
  \includegraphics[width=0.7\textwidth]{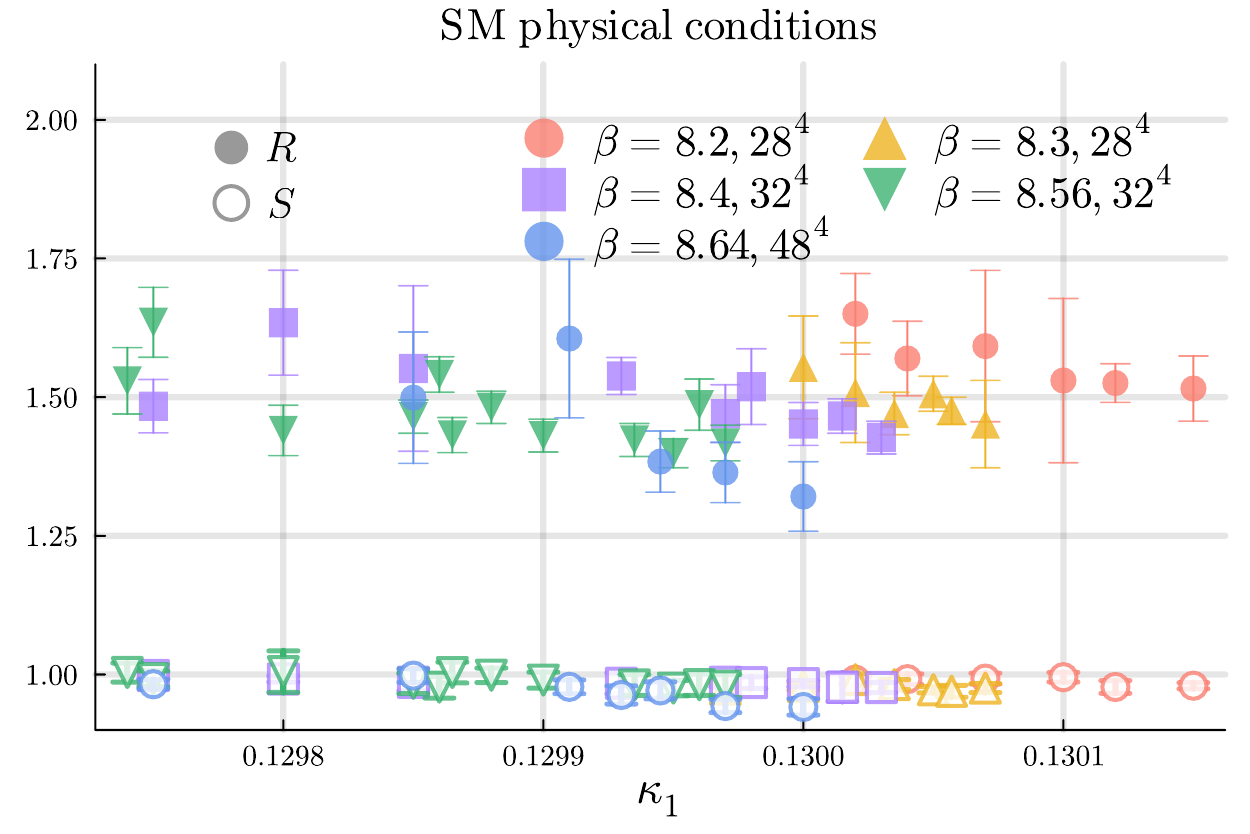}
  \caption{Standard Model conditions \cref{eq:S=1,eq:R=1.5} for simulations at
different $\kappa_{1}$ with other couplings fixed to the values in \cref{tab:LCP}.}
  \label{fig:S_R_BSM_scan}
\end{figure}

Knowing $\kappa_{1}^{c}$ for each $\beta$, we then
perform simulations at different choices of $\kappa_{1}$ around
$\kappa_{1}^{c}$, holding the values of all the other couplings and lattice volumes fixed to those in tab.~\ref{tab:LCP}.  To ensure that we stay on the LPCP when the simulations are carried out in the $(H_{2})$ sector, the SM conditions, eqs.~(\ref{eq:R=1.5}) and (\ref{eq:S=1}), are monitored.  Figure~\ref{fig:S_R_BSM_scan} shows the results of these conditions from the scan of $\kappa_{1}$.  From this plot, we see that the values of $R$ and $S$ remain compatible with those for the LPCP tuned in sec.~\ref{sec:results_tuning}.  In other words, taking into account the uncertainty of
$\kappa_{1}^{c}$ exhibited in~\cref{tab:k1_crit}, one concludes that the SM conditions are still satisfied within error in sector $(H_{2})$ for the values of $\kappa_{1}$ we investigated.  This can be qualitatively understood as a property of the model in the regime of weak quartic couplings, and the current study is indeed performed in this regime.  Nevertheless, we note 
that the LPCP in \cref{tab:LCP} is defined with $\kappa_{1}$ far away from
$\kappa_{1}^{c}$, while the points shown in
\cref{fig:S_R_BSM_scan} are around the transition point.  It is then interesting to see that the SM conditions remain valid with such a large variation of $\kappa_{1}$.  Finally, for
$\kappa>\kappa_{1}^{c}$ the theory is in sector $(H_{12})$, and the SM conditions are not expected to be valid.

\begin{figure}[htb!]
  \centering
  \includegraphics[width=0.49\textwidth]{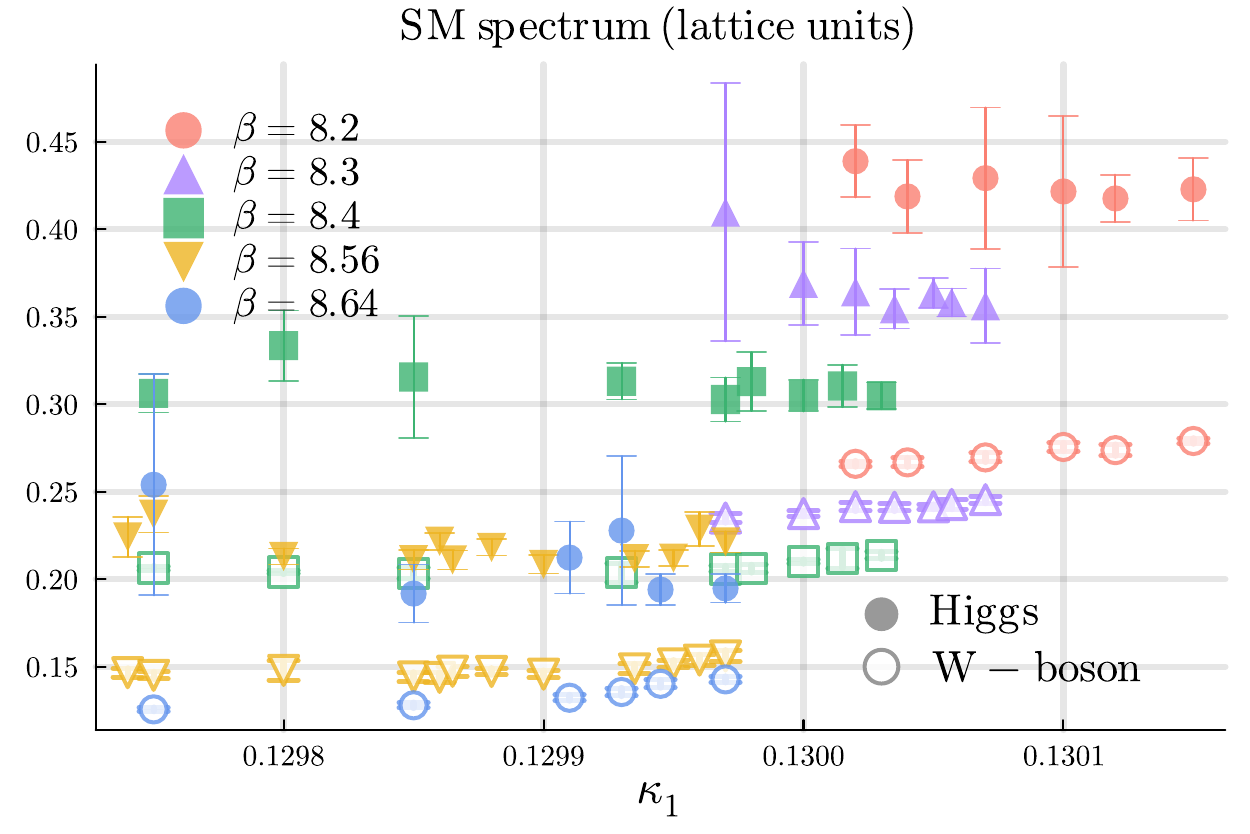}
  \includegraphics[width=0.49\textwidth]{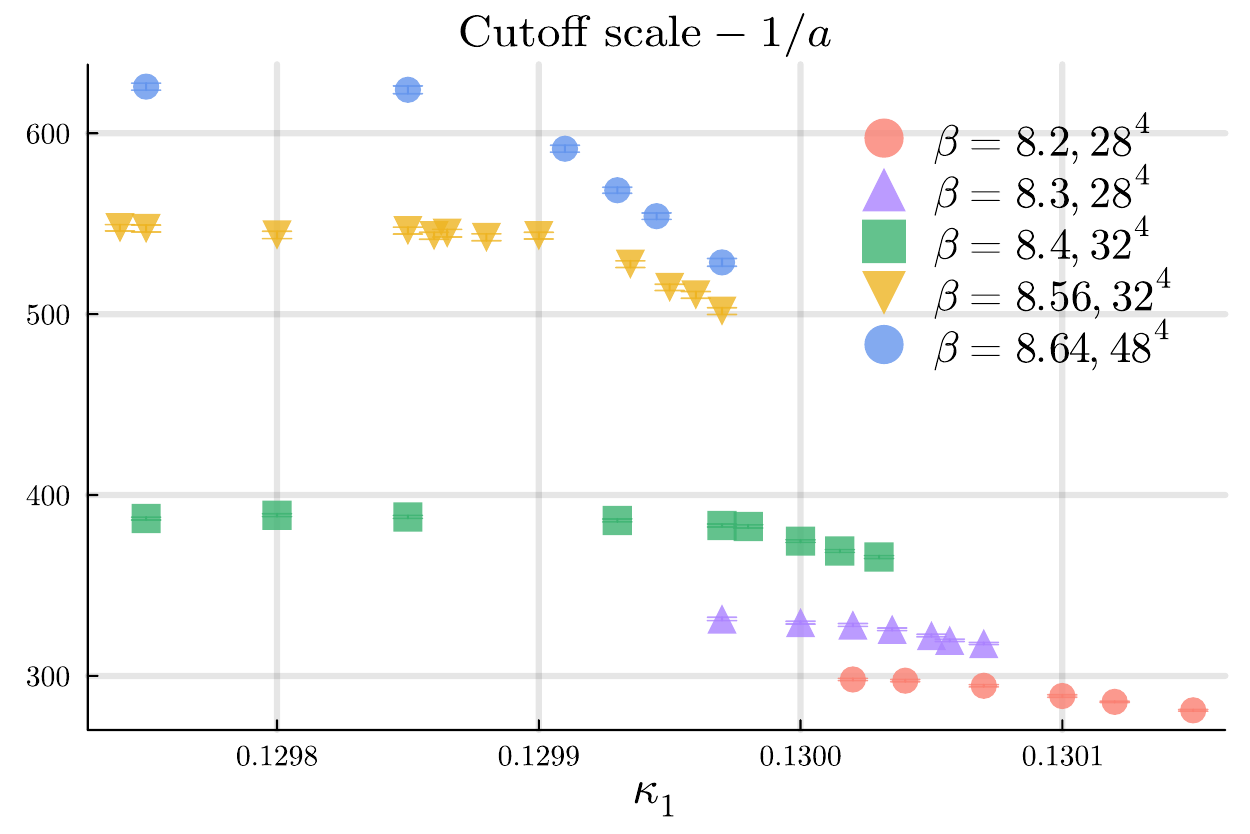}
\caption{Left: Higgs and $W$ boson masses in lattice units as a function of
$\kappa_{1}$ corresponding to the simulations in \cref{fig:S_R_BSM_scan}. Right:
Cutoff energy estimated from the condition
$S=\sqrt{8t_{0}}m_{W}=1$, with $m_W$ set to the experimental value.}
  \label{fig:H_W_cutoff_BSM_scan}
\end{figure}

In addition to $R$ and $S$, we also examine the masses of the Higgs and $W$ bosons in the above scan of $\kappa_{1}$.  We exhibit these results in the left plot of \cref{fig:H_W_cutoff_BSM_scan}, where the $\kappa_{1}$ values are those appearing in \cref{fig:S_R_BSM_scan}.  As can be seen in this plot, these masses in lattice units are compatible with those on the LPCP in tab.~\ref{tab:LCP}.   We notice that at the largest $\kappa_{1}$ for each bare gauge coupling, $am_{h}$ and $am_{W}$ increase slightly, while their ratio, $R$, remains consistent with that at other $\kappa_{1}$ values in the scan.  To investigate this phenomenon further, we study the $\kappa_{1}$ dependence of the cutoff scale. The right plot of
\cref{fig:H_W_cutoff_BSM_scan} shows the cutoff scale, $1/a$,
estimated from the condition $S=\sqrt{8t_{0}}m_{W}=1$ with $m_{W}$
set to its physical value.  This plot demonstrates that the cutoff 
decreases when $\kappa_{1}$ reaches large enough values, and indicates that the increase in $am_{h}$ and $am_{W}$ derive from the change in the lattice spacing.  
Notice, however, that
due to the uncertainty in $\kappa_{1}^{c}$, these points with a decreasing cutoff scale may 
already be in sector $(H_{12})$, where one cannot embed the SM in the model.
\footnote{The improved estimate of $\kappa_{1}^{c}$ for
$\beta=8.56$ in \cref{sec:results_spontaneous_sym_breaking} further confirms
this.}

The key conclusion is that the SM conditions, and thus the LPCP, remain unchanged within sector $(H_{2})$ when $\kappa_1$ is varied.  As will be described in detail below, the masses of the BSM scalar states, $m_{H,A,H^{\pm}}$, decrease by roughly one order of magnitude to become smaller than $m_{W}$ when $\kappa_1$ is varied from 0.1245 (its value in tab.~\ref{tab:LCP}) to $\sim \kappa_{1}^{c}$.
This leads to the expectation that the BSM dynamics of the model in the region near $\kappa_{1}^{c}$ can be quite different from that in the regime wherethe LPCP tuning is carried out, but it is interesting to see that due to the weakness of the quartic coupling, the SM conditions are still valid at $\kappa_{1} \lesssim \kappa_{1}^{c}$.

\begin{figure}[htb!]
\centering
  \includegraphics[width=0.7\textwidth]{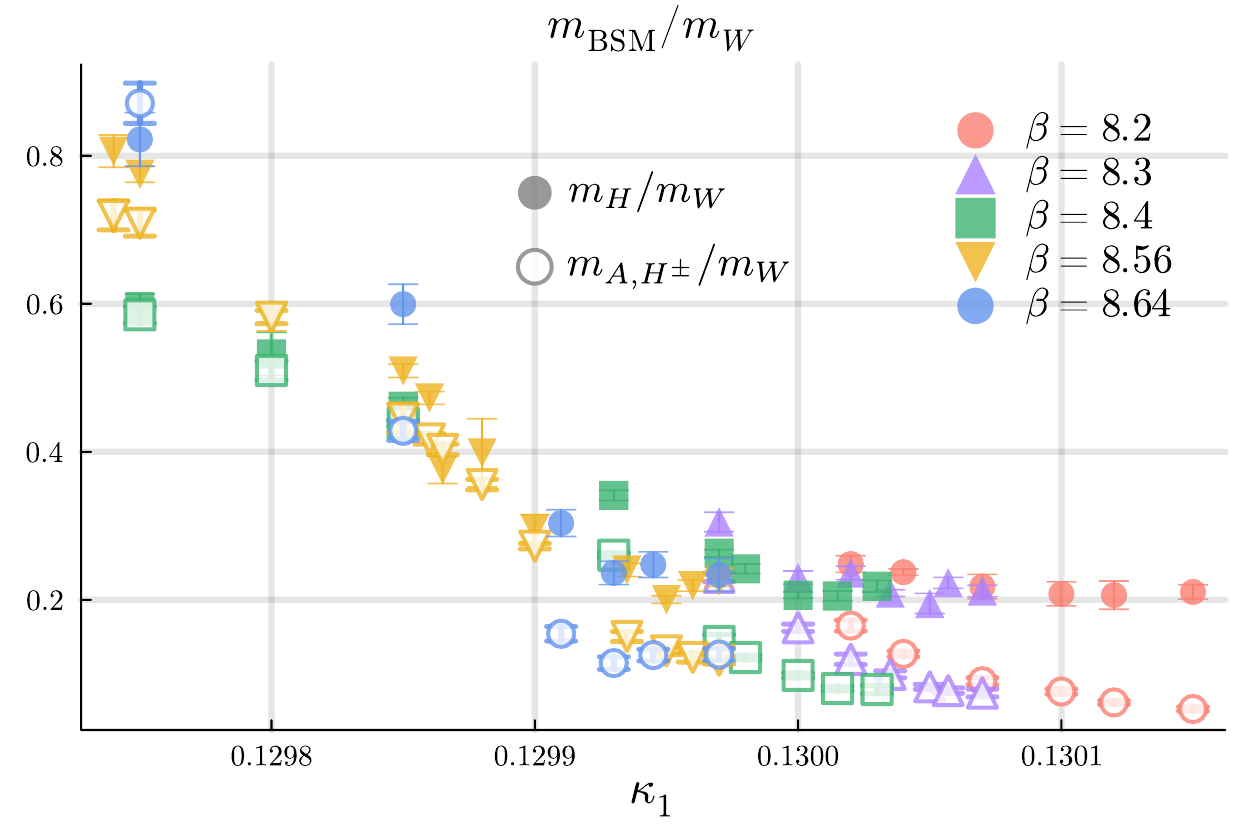}
  \caption{Mass ratio between the BSM states,
$m_{S_{12}^{4}}=m_{H},~m_{S_{12}^{j}}=m_{A}=m_{H^{\pm}}$, and the $W$ boson,
$m_{W}$, for the points in \cref{fig:S_R_BSM_scan} at different $\kappa_{1}$
below $\kappa_{1}^{c}$.}
  \label{fig:mBSMoW_BSM_scan}
\end{figure}

Next, we investigate the spectrum of the BSM scalar states for all $\kappa_{1}{-}$scanning points in \cref{fig:S_R_BSM_scan}. 
Figure~\ref{fig:mBSMoW_BSM_scan} shows the ratios between the masses of these states and $m_{W}$ for all values of $\beta$ in tab.~\ref{tab:LCP}.  Here we stress again that the values of $\kappa_{1}$ in this figure are close to
$\kappa_{1}^{c}$, and are much larger than that ($\kappa_{1}=0.1245$) used to tune the
LPCP in \cref{tab:LCP}.  Comparing the results exhibited in figs.~\ref{fig:lcp} and \ref{fig:mBSMoW_BSM_scan}, we see significant changes in the BSM scalar masses when varying $\kappa_{1}$.  It is also obvious that the mass difference between the $H$ and $A$, $H^{\pm}$ is clearly observed when $\kappa_{1}$ is close to $\kappa_{1}^{c}$, while these four states appear to be degenerate within error in~\cref{fig:lcp}.   Since all the bare quartic couplings are small in this work, one can understand this qualitatively from the tree-level perturbative result in~\cref{eq:tl_mass_matrix}, which demonstrates that this mass difference can be attributed to the additive effect of a non-vanishing $\eta_{4}$.  Being an additive effect, it is then expected to be more pronounced when the BSM scalar masses are smaller.  It should also be mentioned that the data in~\cref{fig:mBSMoW_BSM_scan} are obtained at a higher precision compared to those in~\cref{fig:lcp}. 

One salient feature in~\cref{fig:mBSMoW_BSM_scan} is that $m_{H}/m_{W}$ develops a plateau when $\kappa_{1}$ approaches $\kappa_{1}^{c}$ in sector $(H_{2})$.  This offers evidence for a non-zero lower bound for $m_{H}$ in this sector.  We estimate this lower bound for each $\beta$ value by fitting the data points near $\kappa_{1}^{c}$ in~\cref{fig:mBSMoW_BSM_scan} to a constant.  Results of these fits are displayed in tab.~\ref{tab:mass_lower_bound}.
Contrary to the behaviour of $m_{H}/m_{W}$, it can be seen that $m_{A,H^{\pm}}$ continues to decrease with increasing $\kappa_{1}$.  This is expected since these states become the Goldstone
bosons in sector $(H_{12})$, in the regime $\kappa_{1} > \kappa_{1}^{c}$.  It is also observed that while the plateau value of $m_{H}$ seems
cutoff-independent, the masses of the other three states
increase with the cutoff near the transition.

\begin{table}[htb!]
\centering
\begin{tabular}{cccccc}
    \toprule
 $\beta$              &        8.2 &        8.3 &        8.4 &       8.56 &       8.64 \\
    \midrule
 $m_{H}/m_W$          &  0.209(12) & 0.2115(43) & 0.2080(37) & 0.2111(39) & 0.2271(83) \\
    \bottomrule
\end{tabular}
\caption{Estimated lower bound for the mass ratio of
the BSM $H$ state to the $W$ boson, $m_{H}/m_{W}$, with the values of the quartic couplings and $\kappa_{2}$ fixed to those in tab.~\ref{tab:LCP}.}
\label{tab:mass_lower_bound}
\end{table}


\begin{figure}[htb!]
  \centering
  \includegraphics[width=0.7\textwidth]{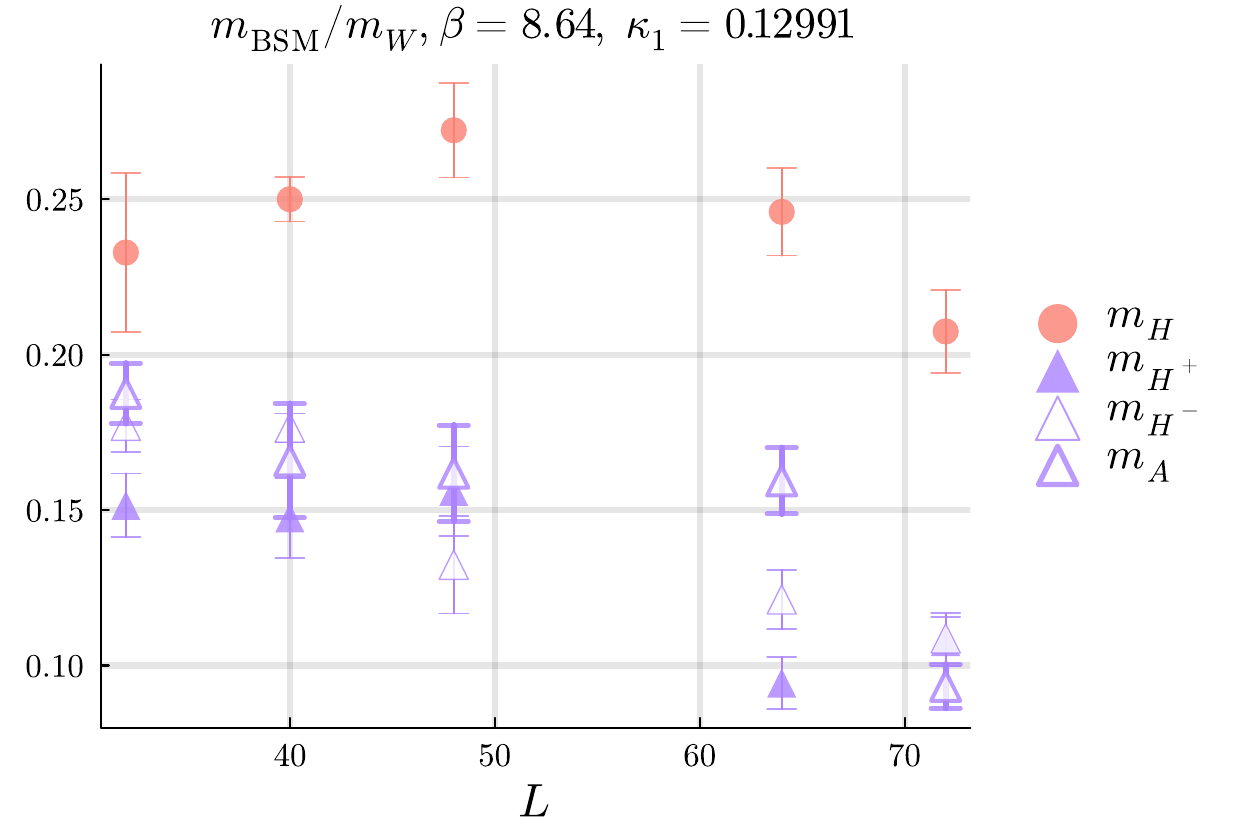}
\caption{Lattice-size dependence of $m_{\rm BSM}/m_{W}$
at the $\beta=8.64$ and $\kappa_{1}=0.12991$, with all other couplings fixed to their LPCP values in tab.~\ref{tab:LCP}). The three degenerate
scalars (triangles) are displayed separately.}
  \label{fig:FV_effects_BSM}
\end{figure}

In the regime where $\kappa_{1}$ is near $\kappa_{1}^{c}$ in sector $(H_{2})$, the BSM scalar states can be much lighter than the SM particles.  Therefore, it is important to check that finite-volume effects are under control.  We illustrate this using the case of $\beta=8.64$.  For this lattice bare gauge coupling, the LPCP discussed in sec.~\ref{sec:results_tuning} is tuned using the lattice size $L=48$, as shown in tab.~\ref{tab:LCP}, where we also report $m_{W} L \sim 6$.  This lattice size should be enough to keep finite-size effects in check for the case in tab.~\ref{tab:LCP}, where the $W$ boson is the lightest state in the spectrum, as indicated by~\cref{fig:lcp}.  However, for $\kappa_1$ approaching $\kappa_{1}^{c}$ in~\cref{fig:mBSMoW_BSM_scan}, the BSM scalar states are lighter than the $W$ boson, while the lattice size is also chosen to be the same as that in tab.~\ref{tab:LCP}.  For the case of $\beta=8.64$ and $\kappa_{1}=0.12991$, the mass of the lightest BSM state is about $15\%$ of $m_{W}$, resulting in a small $m_{\rm BSM} L \sim 1 < m_{W} L$.  Using the results at this $\kappa_{1}$ value to investigate the finite-size effects, simulations at four other lattice volumes, $L=32, 40, 64, 72$, are performed.   Figure~\ref{fig:FV_effects_BSM} shows the volume dependence of $m_{\rm BSM}/m_{W}$ at $\beta=8.64$ and $\kappa_{1}=0.12991$.  It can be observed that $m_{H}/m_{W}$ exhibits mild dependence on $L$, indicating that, at least within errors, volume effects are under control for this mass ratio.  On the other hand, the situation is different for the states $A$ and $H^{\pm}$, where their masses show
clear volume dependence.\footnote{Notice that three masses should be degenerate within statistical error, but only in the infinite-volume limit.  Finite-size effects can break this degeneracy.}
In sector $(H_{2})$ these states
are massive, and the expected volume dependence is exponential
$\order{e^{-mL}}$.  Indeed, infinite-volume extrapolations are investigated using a massive-like exponential function or a massless-like
power law \cite{Hasenfratz:1989ux}.  
Both procedures lead to non-vanishing but small $m_{A}=m_{H^{\pm}}$ in the infinite-volume limit in sector $(H_{2})$ when $\kappa_{1}$ is close to $\kappa_{1}^{2}$.  
This is exactly what is expected, since only in $(H_{12})$
these three states become Goldstone bosons.


%

%
\subsection{Finite temperature transition}
\label{sec:results_finite_temp}

In this subsection we investigate the custodial 2HDM at finite
temperature on the LPCP tuned in~\cref{sec:results_tuning}.  In order to study the temperature dependence we perform
simulations using the bare couplings in \cref{tab:LCP} and asymmetric lattices with sizes $L^{3}\times L_{4}$, where
the temporal extent, $L_{4}$, defines the physical temperature $T=1/(aL_{4})$.
The lattice sizes at the corresponding $\beta$ values are
summarized in tab.~\ref{tab:simulations_fin_temp}.

\begin{table}[htb!]
  \centering
  \begin{tabular}{ccc}
    \toprule
    $\beta$ & $L$ & $L_{4}$   \\
     \midrule
8.2 & 28 & 2,3,4,5,6,7,8 \\
8.3 & 28 & 2,3,4,5,6,7,8  \\
8.4 & 32 & 2,3,4,5,6,7,8,10  \\
8.56 & 32 & 2,3,4,5,6,7,8,10,12,14,16  \\
8.64 & 48 & 2,3,4,5,6,7,8,10,12,16,20  \\
    \bottomrule
  \end{tabular}
  \caption{Finite temperature simulations for each $\beta$ value in
the LPCP. The spatial extent $L$ and temporal extent $L_{4}$ are shown
for each $\beta$. The values of the remaining couplings are listed in
\cref{tab:LCP}.  }
  \label{tab:simulations_fin_temp}
\end{table}


Since we are interested in studying the `symmetry restoration' that deactivates
the Higgs/FMS mechanism, we examine the observables $\rho^{2}_{2}$ and $L_{\alpha_{2}}$ defined in eqs.~(\ref{eq:rho_def}) and (\ref{eq:global_obs}), which signal the
transition from $(H_{2})$ to $(H_{0})$.
For each value of $\beta$ in the LPCP
(\cref{tab:LCP}), simulations at 
different time extents can be employed to probe temperature-dependent 
properties of these two observables, generically denoted as $\mathcal{O}(T)$ in this subsection.  Contrary to the masses obtained from the lattice correlation functions,
quantities such as $\rho_{2}^{2}$ or $L_{\alpha_{2}}$ need to be renormalized.
Since the exact renormalization of $\mathcal{O}$ is beyond the scope of this work, we examine the ratio, $\langle {\mathcal{O}}(T)\rangle/\langle {\mathcal{O}}(0) \rangle$, in which the multiplicative renormalization of the operator cancels.
This quantity can then be compared for different $\beta$, with possible differences being due to lattice artifacts and effects arising from the mistuning and the mismatch of the BSM spectrum in the LPCP.

\begin{figure}[htb!]
  \centering
  \includegraphics[width=0.49\textwidth]{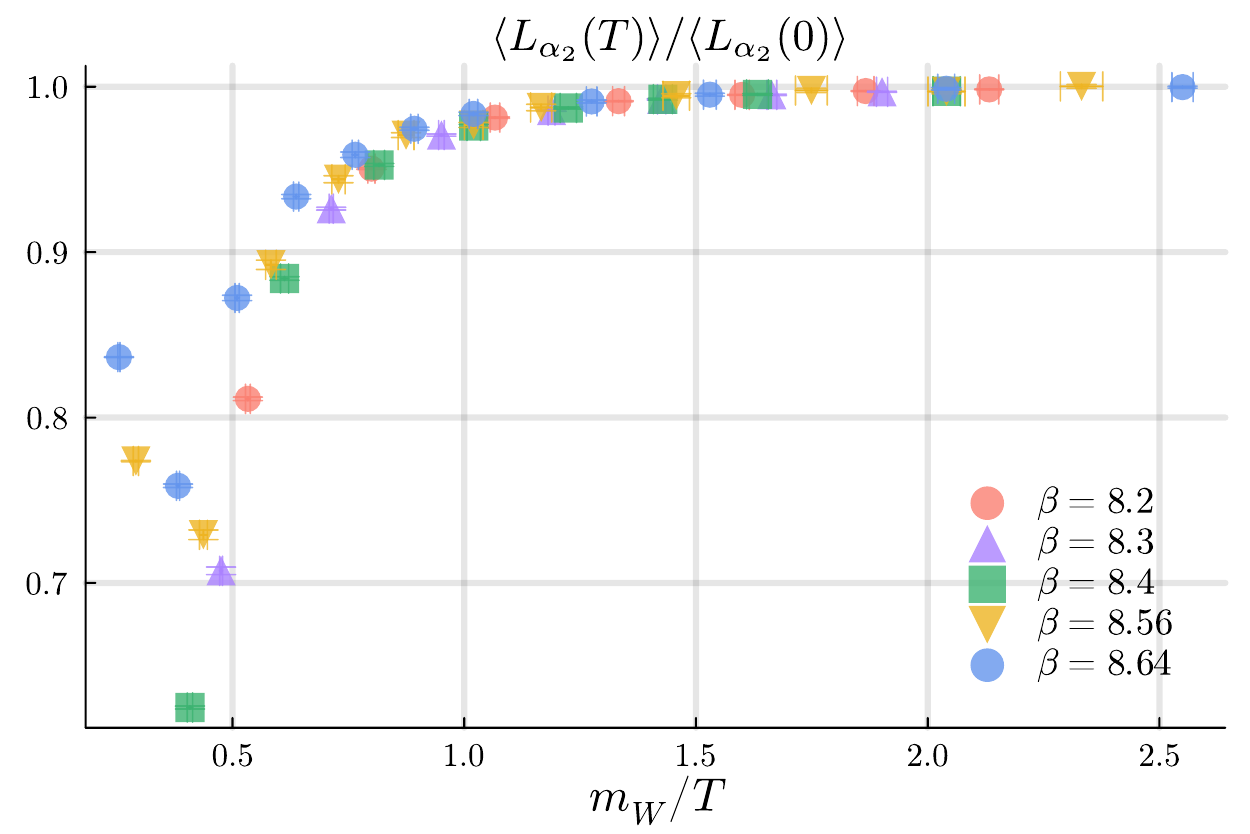}
  \includegraphics[width=0.49\textwidth]{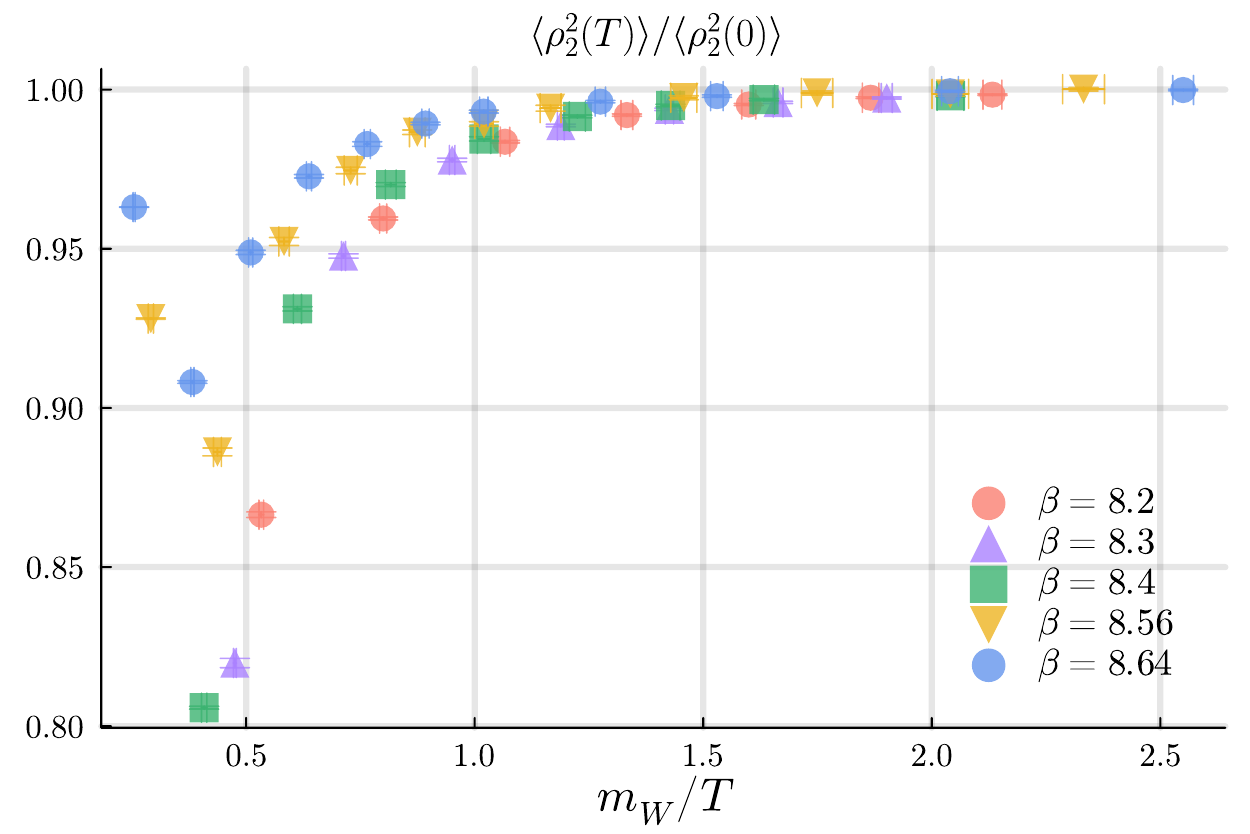}
\caption{Finite-temperature dependence of $\langle L_{\alpha_{2}}(T)\rangle/\langle L_{\alpha_{2}}(0)\rangle$ (left) and
$\langle \rho_{2}^{2}(T)\rangle/\langle \rho_{2}^{2}(0)\rangle$ (right) on the LPCP at different $\beta$ values in tab.~\ref{tab:LCP}.}
  \label{fig:ratio_vev_fintemp}
\end{figure}

Figure~\ref{fig:ratio_vev_fintemp} exhibits the ratios $\langle L_{\alpha_{2}}(T)\rangle/\langle L_{\alpha_{2}}(0)\rangle$ and $\langle\rho_{2}^{2}(T)\rangle/\langle\rho_{2}^{2}(0)\rangle$ as functions of the dimensionless quantity
$ m_{W}/T$, with $am_{W}$ taken from its zero-temperature values in 
tab.~\ref{tab:LCP}.  Notice that since the temperature
increases with the approach to the continuum at fixed $L_{4}$, larger values of $\beta$
provide access to larger temperatures.
In the left plot, results for $\langle L_{\alpha_{2}}(T)\rangle/\langle L_{\alpha_{2}}(0)\rangle$ at various $\beta$ values show small or moderate differences in the regime $T < m_{W}$.  At high-temperature, this difference becomes more sizable since the simulations are carried out at small $L_{4}$, resulting in large lattice artefacts whose effects are particularly significant at $L_{4}=2$.  On the other hand, $\langle\rho_{2}^{2}(T)\rangle/\langle\rho_{2}^{2}(0)\rangle$, as displayed in the right plot, shows much more significant discrepancies at different lattice spacings.  This can be understood by the fact that $\rho_{2}^{2}$ still needs to be additively renormalized, while $L_{\alpha_{2}}$ does not.  For this reason, below we will only use $L_{\alpha_{2}}(T)$ to study the phase transition.

\begin{figure}[htb!]
  \centering
  \includegraphics[width=0.49\textwidth]{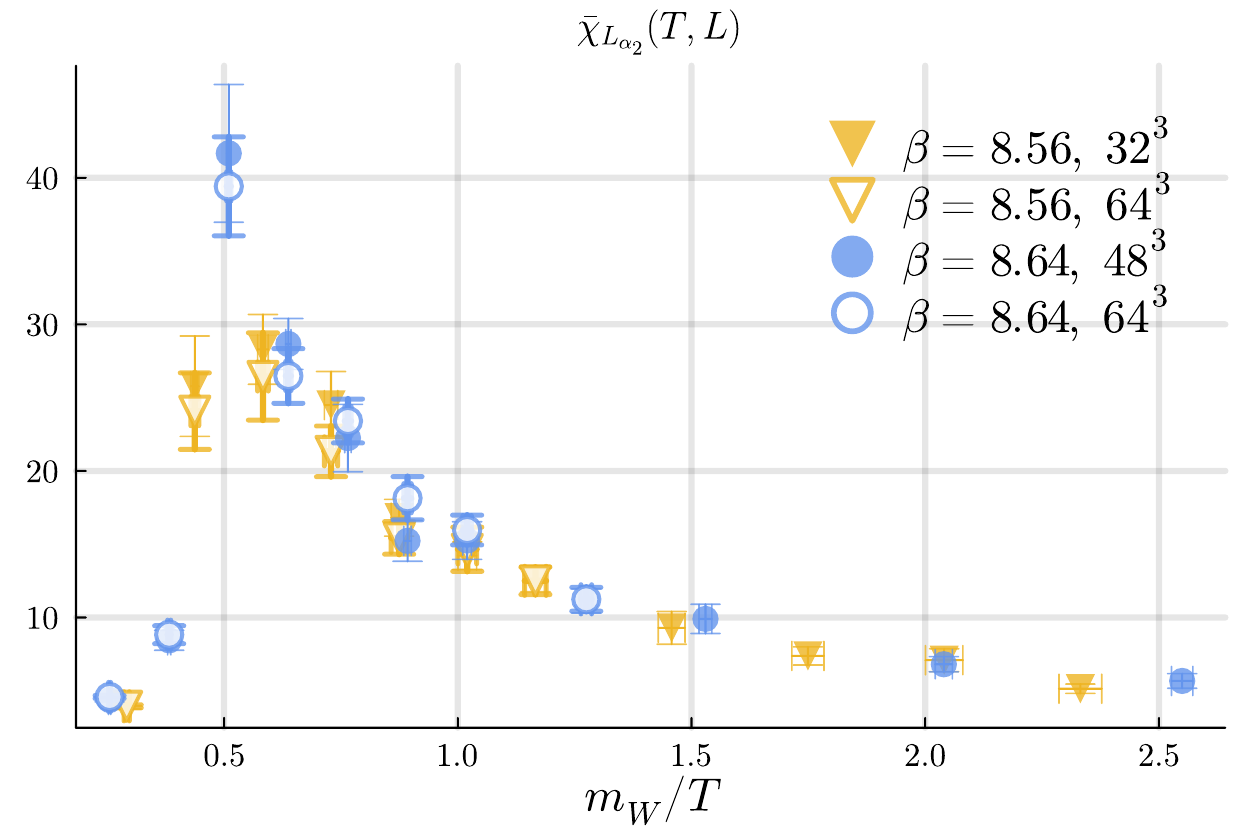}
\caption{Finite-temperature dependence of the ratio of the normalized susceptibility,
$\bar\chi_{L_{\alpha_{2}}}$, defined in eq.~(\ref{eq:norm_susceptibility}). Data at two different spatial volumes are shown for each of the two largest $\beta$ values.}
  \label{fig:sus_lcp_fintemp}
\end{figure}

From~\cref{fig:ratio_vev_fintemp}, one can conclude that a finite-temperature transition occurs around $m_{W}/T_{c} \sim 0.5$, where $T_{c}$ is the critical temperature.  In order to better resolve the transition point, and to
understand the character of this phase transition we resort to the susceptibility of $L_{\alpha_{2}}$,
\begin{equation}
  \label{eq:susceptibility}
  \chi_{L_{\alpha_{2}}}(T,L) = L^{4}\left( \langle L_{\alpha_{2}}^{2}(T,L) \rangle - \langle L_{\alpha_{2}}(T,L)\rangle^{2}\right)\, ,
\end{equation}
at various temperatures and lattice volumes.  In the above definition, we make the volume dependence in $L_{\alpha_{2}}(T)$ explicit by denoting it as $L_{\alpha_{2}} (T,L)$. Again, to avoid dealing with the issue of renormalizing the operator $L_{\alpha_{2}}$, we compute the "normalized susceptibility" defined as
\begin{equation}
  \label{eq:norm_susceptibility}
  \bar{\chi}_{L_{\alpha_{2}}}(T,L) \equiv \frac{\chi_{L_{\alpha_{2}}}(T,L)}{\chi_{L_{\alpha_{2}}}(0,L)} =\frac{\langle L_{\alpha_{2}}^{2}(T,L) \rangle - \langle L_{\alpha_{2}}(T,L)\rangle^{2}}{\langle L_{\alpha_{2}}^{2}(0,L) \rangle - \langle L_{\alpha_{2}}(0,L)\rangle^{2}} \, .
\end{equation}
Figure~\ref{fig:sus_lcp_fintemp} demonstrates the temperature dependence of $\bar{\chi}_{L_{\alpha_{2}}}(T,L)$. In this plot, we only show results from the two largest $\beta$ values in tab.~\ref{tab:simulations_fin_temp}.  This is because for coarser lattices, either the data points in the regime of $m_{W}/T \sim 0.5$ are expected to suffer from significant effects of lattice artefacts ($\beta=8.4$), or there are no data points in this transition regime ($\beta=8.2$ and 8.3).   For each of the $\beta$ value shown in this figure, we perform an additional simulation at a larger spatial volume.

Results in~\cref{fig:sus_lcp_fintemp} indicate that the critical temperature (the location of the peak for $\bar{\chi}_{L_{\alpha_{2}}} (T,L)$ as a function of $m_{W}/T$) exhibits mild lattice-spacing dependence.  The same behavior was also found previously on
a lattice study for the single-Higgs-doublet model~\cite{D_Onofrio_2016}.  In any case, our data demonstrate that the critical temperature is approximately
\begin{equation}
\label{eq:T_c}
 T_{c} \sim \frac{m_{W}}{0.5} \approx 160 \mbox{ }\mathrm{GeV} \, .
\end{equation}
We stress that the purpose of this work is not to determine $T_{c}$, and the result in the above equation should be regarded as a rough estimate.  Instead, the relevant information from fig.~\ref{fig:sus_lcp_fintemp} is the absence of significant volume dependence in the susceptibility.  This strongly implies that the transition is a crossover.  
Our finding for the finite-temperature transition in this custodial 2HDM at weak quartic couplings is then compatible with that in the SM, where previous lattice investigations~\cite{DOnofrio:2014rug,D_Onofrio_2016} concluded for a crossover at 
$T_{c} = 159.5(15)~\si{GeV}$.


Perturbatively, most studies
\cite{Dorsch:2014qja,dorsch_strong_2013,basler_strong_2017,bernon_new_2018}
suggest that large quartic couplings (equivalently, a large mass splitting in the
BSM scalar sector) are needed for the finite-temperature phase transition to be first-order in 2HDM.  In this
sense, although the lattice cutoffs considered in this work are not large enough
to accurately resolve the electroweak transition in the inert 2HDM, our results
provide a non-perturbative confirmation that indeed, small quartic couplings do
not lead to a first-order transition.  There are, however, recent perturbative
predictions~\cite{bernon_new_2018} for a first-order thermal phase transition in the 2HDM for
degenerate BSM scalars with masses around $350~\si{GeV}$.  It turns out that
these conditions for the mass spectrum are satisfied for the two finest lattices in this study (the BSM scalar
masses are shown in \cref{fig:lcp}).  Nevertheless, we observe no sign of first-order transition in our non-perturbative lattice investigation.  

\section{Conclusion and outlook}
\label{sec:conclusion}

This paper presents the first step of our research programme on non-perturbative investigations of the custodial 2HDM using the techniques of lattice field theory.  While our eventual goal is to work in the scenario of strong {\it renormalized} quartic couplings, as a first step, however, we start with the regime where all the quartic couplings are weak.  This strategy enables the initial calibration of the theory by understanding its non-thermal phase structure, and facilitates the study of the theory without dealing with any triviality issues.  It also allows us to set the stage for the exploration of the following three crucial issues.  First, for the custodial 2HDM to be realized in nature, it has to be able to reproduce the low energy SM observations. Second, it is of interests to the community of particle physicists to gain knowledge on the bounds of the masses of the BSM scalar states.  Last, it is essential to understand whether there is a strong first-order EWPT in 2HDM. Althought there have been many perturbative studies of the above issues, a non-perturbative, first-principle investigation is still desirable.

In the work reported here, we consider the parameter space of the custodial 2HDM that can be divided in four sectors, as summarized in fig.~\ref{fig:scheme_sectors}. Similarly to the case of the SM, most of these regions can be connected through crossover transitions, and therefore do not define truly different phases of the model. 
Among the four,  $(H_{0})$
corresponds to the confining region, while the Higgs/FMS mechanism is active in $(H_{1})$, $(H_{2})$ and
$(H_{12})$.   
Contrary to the SM, the additional degrees of freedom in this model allow for spontaneous symmetry breaking to occur. In order to probe this and understand in detail the properties of the global symmetries, we examine the order parameter, $\langle L^{3}_{\alpha_{12}}\rangle$, defined in eq.~(\ref{eq:global_obs}).  This order parameter can be employed to probe the $SU(2)$ custodial symmetry, which is shown to be spontaneously broken in $(H_{12})$ through our careful study described in sec.~\ref{sec:results_spontaneous_sym_breaking}.  This means that if we want to keep the $SU(2)$ custodial symmetry, only sectors $(H_{1})$ and $(H_{2})$ are viable for model building purposes.  That is, it is possible to embed the SM in the custodial 2HDM only in these two sectors.  Since $(H_{1})$ and $(H_{2})$ are exactly the same modulo naming conventions, we choose to work in $(H_{2})$ in this project.

To study the embedding of the SM in the $(H_{2})$ sector of the custodial 2HDM, we consider a `line of partial constant physics', acronymized as LPCP throughout this article, that is defined to be a line in the bare parameter space, on which the SM physics is realized at different (high enough) cutoff scales. 
 To be more precise, two conditions involving only SM particles and interactions are being kept.  The first is the ratio between the Higgs and $W$ boson masses, and the second is the value of the renormalized gauge coupling.  These conditions are given in eqs.~(\ref{eq:R=1.5}) and (\ref{eq:S=1}).  
 The first non-trivial task in our non-perturbative investigation of the 2HDM is to confirm that such a LPCP does exist.  Working with the scenario of weak quartic couplings makes the task simpler, and it is why we choose to perform our first study in this regime.  In sec.~\ref{sec:results_tuning}, we show that this can be achieved by adjusting the bare gauge coupling, as well as the other two couplings that involve only the SM Higgs doublet, namely, $\kappa_{2}$ and $\eta_{2}$ in the action in eq.~(\ref{eq:lattice action}).  All the other bare couplings (the BSM couplings) are kept fixed in this tuning procedure.  It is demonstrated that we can obtain a LPCP for five different cutoff scales straddling between 300 and 630 GeV.  Details of this LPCP are listed in tab.~\ref{tab:LCP}.

For the tuning of LPCP, we employ the gradient-flow scheme to define the renormalized gauge coupling, $g^{2}_{GF}$.  Interesting features are expected for the running behaviour of the coupling in this mass-dependent scheme, since we work in the Higgs phase $(H_{2})$ where the gauge bosons are massive. Section~\ref{sec:results_gauge_coupling} illustrates our exploration of this issue.  As shown in fig.~\ref{fig:running_coupling}, at renormalization scales that are much larger than $m_{W}$, $g^{2}_{GF}$ behaves in a similar way as the QCD coupling, exhibiting asymptotic freedom.  Its running can be well described by the mass-independent one-loop $\beta{-}$function given in eq.~(\ref{eq:betafunction}).  At energy scales around $2.5m_{W}$, where effects from the screening of the gauge force due to the $W$ boson mass are relevant, $g^{2}_{GF}$ starts to deviate from the one-loop mass-independent running behaviour.   In the regime where the renormalization scale is below $m_{W}$, the characteristics of its energy dependence can be captured by a Yukawa-like potential.  We stress that our work is the first time the gradient-flow gauge coupling is studied in a Higgs phase of gauge-scalar theories.

The aforementioned LPCP is obtained at fixed bare BSM couplings, which results in the variation of the BSM spectrum with the cutoff scale, as can be seen in fig.~\ref{fig:lcp}.  Since no BSM particles have been found in experiments hitherto, it is not useful to try to obtain a complete line of constant physics for the 2HDM by also keeping the masses of the BSM states fixed.  Instead, it is more interesting to change the BSM couplings and investigate bounds on the additional scalar  masses that are compatible with the SM conditions.  For this purpose, we perform a scan in the hopping parameter ($\kappa_{1}$) for the BSM Higgs doublet.  In sec.~\ref{sec:results_spectrum}, we give details of this aspect of our work.  Our strategy is to vary $\kappa_{1}$ within the $(H_{2})$ sector, while keeping all the other couplings fixed to their values used to tune the LPCP.  It is found that the model still stays on the LPCP in this procedure.  This can be understood by the nature of the weak quartic couplings.  The most important conclusion of this scanning is that the BSM scalar states can be as light as about $0.15 m_{W}$, and almost as heavy as the cutoff scale.  Notice also that from experimental searches, the existence of BSM scalar particles with very low masses is not ruled out.  There are recent works in this direction
\cite{CMS:2019hvr, ATLAS:2024bjr,CMS:2024yhz} and proposals to improve the studies at
energies below 100~GeV. 
In \cite{CMS:2024yhz} the range $25-70~\textrm{GeV}$ is
probed for the pseudoscalar $m_{A}$ by looking for $\tau^{-}\tau^{+}$ resonances, while
in \cite{ATLAS:2024bjr} the search focuses on the diphoton final states around the range $70-110~\textrm{GeV}$.

We also study the finite temperature phase transition in this work, by performing simulations at different temporal extents with all the other parameters fixed to their values on the LPCP.  From the analysis  detailed in sec.~\ref{sec:results_finite_temp}, it is concluded that the custodial 2HDM with our choice of weak quartic couplings does not provide a first-order EWPT.  Instead, through carrying out finite-size scaling for the susceptibility of the order parameter, it can be shown that the finite-temperature transition in $(H_{2})$ sector of the theory is a crossover that occurs at around the temperature $T \sim 2 m_{W}$.  This is similar to the scenario of the SM, and is in contrast with the prediction for a first-order phase transition from the perturbative investigation in~\cite{bernon_new_2018}.

The above findings for the bounds on the BSM masses, as well as the finite-temperature crossover behaviour, are from simulations performed at weak quartic couplings.  Working in this weak-coupling regime can make our task simpler.  This is because the interaction between the SM and the BSM sectors is small, such that the LPCP becomes largely insensitive to the change of some bare parameters (such as $\kappa_{1}$ and $\eta_{1}$ in eq.~(\ref{eq:lattice action})) that numerically only affect the BSM sector.  It is a suitable choice for the first step in our research programme, from which we have gained a significant amount of experience in tuning and calibrating the lattice computations, and have found interesting features of the custodial 2HDM.  While this is an
advantage from the practical perspective, there is no fundamental reason for
the choice of small renormalized quartic couplings, and a complete investigation is necessary. In fact, many perturbative studies suggest that 
$\order{1}$ quartic couplings are required for a strong
first-order EWPT.  For instance, tree-level results
\cite{basler_strong_2017,bernon_new_2018} indicate the need of large mass
splitting amongst the BSM scalar states for the occurrence of this
first-order transition.  This condition is directly related to an enhanced value of
the inter-flavour Higgs quartic coupling, namely, $\eta_{3}$.  Our future work will extend this exploration to the case of strong {\it renormalized} quartic couplings, in particular $\eta_{3}$.   The tuning for a LPCP is expected to be much more challenging in this regime due to the renormalization
effects of these couplings.  In fact, given that the model is expected to be trivial, it is not even obvious whether a LPCP can be realized at large enough cutoff scales when some of the renormalized quartic couplings are large. 
To keep the value of a renormalized quartic coupling fixed at $\mathcal{O}(1)$, non-perturbative implementation of a suitable renormalization scheme is required.  For this purpose, the use of the gradient flow on the scalar fields~\cite{Monahan:2015lha}, similar to that on the gauge fields for the definition of $g_{GF}^{2}$, will be useful. Moreover, such a scheme is a prerequisite to investigate the issue of triviality in gauged scalar theories.
To keep the value of a renormalized quartic coupling fixed at $\mathcal{O}(1)$, non-perturbative implementation of a suitable renormalization scheme is required.  For this purpose, the use of the gradient flow on the scalar fields~\cite{Monahan:2015lha}, similar to that on the gauge fields for the definition of $g_{GF}^{2}$, will be useful. Moreover, such a scheme is a prerequisite to investigate the issue of triviality in gauged scalar theories.

\acknowledgments
GC and AR would like to acknowledge  financial support from the Generalitat Valenciana (GENT program CIDEGENT/2019/040), Ministerio de Ciencia e Innovacion (PID2020-113644GB-I00).  GC, AH, WSH, CJDL, AR and MS have received support from the CSIC-NSTC exchange grant 112-2927-I-A49-508. CJDL's work is partially supported by the NSTC project 112-2112-M-A49-021-MY3.  Support from the NSTC project 112-2639-M-002-006-ASP is aknowledged by AH, WSH, CJDL and MS.  This work also received funding from the European Union’s Horizon Europe Framework Programme (HORIZON) under the ERA Chair scheme with grant agreement no.101087126 and the Ministry of Science, Research and Culture of the State of Brandenburg within the Centre for Quantum Technologies and Applications (CQTA). The authors thank ASGC (Academia Sinica Grid Computing Center, AS-CFII-114-A11, NSTC(NSTC 113-2740M-001-007) for provision of computational resources.  Numerical computations for this work were also carried out on the HPC facilities at National Taiwan University and National Yang Ming Chiao Tung University.

\appendix
\section{Hybrid Monte-Carlo simulation for two-Higgs doublet models}
\label{sec:HMC}

We use the HMC algorithm in our simulations.  This algorithm introduces a fictitious
(simulation) time $t$ on which all fields depend.  The dynamics in the
fictitious time is dictated by the Hamilton's equations of the following
Hamiltonian
\begin{align}
  \label{eq:Hamiltonian}
  H(\Phi, U; \pi, p) &= \frac{1}{2}\sum_{n,\alpha}\left(\pi_{n}^{\alpha}\right)^{2} +\frac{1}{2}\sum_{\mu,a}\Tr\left(p_{\mu}^{2}\right)  + S_{\text{2HDM}} \, ,
\end{align}
where $p_{\mu}=p_{\mu}^{a}(x)\sigma_{a},~p_{\mu}^{a}\in\mathbb{R}$ are the
algebra-valued momentum fields associated with the gauge fields and
$\pi_{n}^{\alpha}$ are the real momenta conjugate to the real fields
$\varphi_{n}^{\alpha}$.  The evolution of the fields is governed by
\begin{align}
    &\dot \pi_{n}^{\alpha}(x,t) = -\frac{\delta H}{\delta\varphi_{n}^{\alpha}(x,t)}\, , && \dot p_{\mu}^{a}(x,t) = -\partial_{x,\mu}^{a}H \, , \nonumber \\
   &\dot \varphi_{n}^{\alpha}(x,t) = \frac{\delta H}{\delta\pi_{n}^{\alpha}(x,t)} = \pi_{n}^{\alpha}(x,t)\, , && \dot A_{\mu}^{a}(x,t) =   \frac{\delta H}{\delta p_{\mu}^{a}(x,t)}\, ,
\end{align}
where $\partial_{x,\mu}^{a}$ is the algebra-valued differential operator defined
in appendix A of~\cite{Luscher_2010_flow}, and $A_{\mu}^{a}$ the algebra-valued gauge fields,
$U_{\mu} = {\mathrm{exp}}(-iA_{\mu})$.  The time evolution of $U_{\mu}$ can be written as
\begin{equation}
  \dot U_{\mu}(x,t) = -i p_{\mu}(x,t) U_{\mu}(x,t),~(\text{no sum in}~\mu)\, ,
\end{equation}
while numerically the evolution of the group-valued gauge fields is implemented by
\begin{equation}
  \label{eq:4}
 U_{\mu}(x, t+\delta t) = \exp\{-ip_{\mu}(x,t)\delta t\}U_{\mu}(x,t)\, .
\end{equation}
The corresponding momentum variables satisfy
\begin{align*}
  i\dot p_{\mu}^{a}(x,t) =& -2\sum_{n}k_{n} \Tr \left[ \hat\Phi_{n}^{\dagger}(x)\sigma_{a}U_{\mu}(x)\hat\Phi_{n}(x+\hat\mu) \right]\nonumber \\
  &-2\beta\sum_{\nu} \Tr \left[ \sigma_{a}\left( U_{\mu\nu}(x) -  U_{\mu\nu}^{\dagger}(x) \right) - \sigma_{a}\left( U_{\mu\nu}(x-\hat\nu) -  U_{\mu\nu}^{\dagger}(x-\hat\nu) \right) \right]\, .
\end{align*}

Finally, the equation of motion for the scalar momenta in the quaternion form are
\begin{align}
  -\dot \pi_{1}(y,t) =  &- k_{1}\sum_{\mu}\left\{ U_{\mu}(y)\Phi_{1}(y+\hat\mu) + \Phi_{1}(y-\hat\mu) U_{\mu}(y-\hat\mu) \right\} \nonumber \\
  &+ \Phi_{1} + 4\eta_{1} \left[ \Tr \left( \hat\Phi_{1}^{\dagger}\hat\Phi_{1} \right) - 1 \right] \Phi_{1} + 2\mu^{2}\Phi_{2}+ 2\eta_{3}\Tr \left( \hat\Phi_{2}^{\dagger}\hat\Phi_{2} \right)\Phi_{1} \nonumber\\
  &+ 2\eta_{4}\left[\Tr \left( \hat\Phi_{1}^{\dagger}\hat\Phi_{2} \right) \Phi_{2} +\Tr \left( \hat\Phi_{1}^{\dagger}\hat\Phi_{2} i\sigma_{3}\right)\Phi_{2}i\sigma_{3} \right] \nonumber\\
  &+ 2\eta_{5}\left[\Tr \left( \hat\Phi_{1}^{\dagger}\hat\Phi_{2} \right) \Phi_{2} -\Tr \left( \hat\Phi_{1}^{\dagger}\hat\Phi_{2} i\sigma_{3}\right)\Phi_{2}i\sigma_{3} \right] \nonumber\\
  &+ 2   \bigg[\eta_{6}\Tr \left( \hat\Phi_{1}^{\dagger}\hat\Phi_{1} \right) + \eta_{7}\Tr \left( \hat\Phi_{2}^{\dagger}\hat\Phi_{2} \right) \bigg]\Phi_{2} + 4\eta_{6}\Tr \left( \hat\Phi_{1}^{\dagger}\hat\Phi_{2} \right)\Phi_{1}\, ,
\end{align}
\begin{align}
  -\dot \pi_{2}(y,t) &=  - k_{2}\sum_{\mu}\left\{ U_{\mu}(y)\Phi_{2}(y+\hat\mu) + \Phi_{2}(y-\hat\mu) U_{\mu}(y-\hat\mu) \right\} \nonumber \\
  &+ \Phi_{2} + 4\eta_{2} \left[ \Tr \left( \hat\Phi_{2}^{\dagger}\hat\Phi_{2} \right) - 1 \right] \Phi_{2} + 2\mu^{2}\Phi_{1}+ 2\eta_{3}\Tr \left( \hat\Phi_{1}^{\dagger}\hat\Phi_{1} \right)\Phi_{2} \nonumber\\
  &+ 2\eta_{4}\left[\Tr \left( \hat\Phi_{1}^{\dagger}\hat\Phi_{2} \right) \Phi_{1} -\Tr \left( \hat\Phi_{1}^{\dagger}\hat\Phi_{2} i\sigma_{3}\right)\Phi_{1}i\sigma_{3} \right] \nonumber\\
  &+ 2\eta_{5}\left[\Tr \left( \hat\Phi_{1}^{\dagger}\hat\Phi_{2} \right) \Phi_{1} +\Tr \left( \hat\Phi_{1}^{\dagger}\hat\Phi_{2} i\sigma_{3}\right)\Phi_{1}i\sigma_{3} \right] \nonumber\\
  &+ 2   \bigg[\eta_{6}\Tr \left( \hat\Phi_{1}^{\dagger}\hat\Phi_{1} \right) + \eta_{7}\Tr \left( \hat\Phi_{2}^{\dagger}\hat\Phi_{1} \right) \bigg]\Phi_{2} + 4\eta_{7}\Tr \left( \hat\Phi_{1}^{\dagger}\hat\Phi_{2} \right)\Phi_{2} \, .
\end{align}

The numerical implementation of multi-Higgs theories for GPU is available in \cite{lgpu_su2}.  Other than
the HMC implementation, the code contains the
computation of scalar observables, multi-Higgs and vector interpolators, smearing of gauge and scalar fields, as well as the
gradient-flow tools such as the numerical solvers (including an adaptive
step-size integrator) measurement functions for the action density, topological charge, etc.

%


\bibliographystyle{apsrev.bst} 
\bibliography{refs} 
 
\end{document}